\documentclass[12pt]{article}
\usepackage{amsmath,amssymb,exscale}
\usepackage{graphicx}
\usepackage{epsfig}
\usepackage{multicol}
\usepackage{color}
\usepackage{mathrsfs}
\usepackage{blindtext}
 \usepackage{fancyhdr}
\usepackage{hyperref}
\usepackage{cite}
\usepackage{mathtools}
\usepackage{setspace}

\numberwithin{equation}{section}

\usepackage{xcolor}
\usepackage{sectsty}

\usepackage{mdframed}
\usepackage{titletoc}

\definecolor{secnum}{RGB}{13,151,225}
\definecolor{ptcbackground}{RGB}{212,237,252}
\definecolor{ptctitle}{RGB}{0,177,235}

\titlecontents{lsection}
  [5.8em]{\sffamily}
  {\color{secnum}\contentslabel{2.3em}\normalcolor}{}
  {\titlerule*[1000pc]{.}\contentspage\\\hspace*{-5.8em}\vspace*{5pt}%
    \color{white}\rule{\dimexpr\textwidth-15.5pt\relax}{1pt}}

\usepackage{hyperref}
\hypersetup{colorlinks,bookmarksopen,bookmarksnumbered,citecolor=rossos,
linkcolor=redy,pdfstartview=FitH,urlcolor=green-go}
\usepackage{slashed}

\definecolor{blus}{cmyk}{1,1,0,0.1}
\definecolor{verdes}{cmyk}{0.99,0,0.59,0.65}
\definecolor{rossos}{cmyk}{0,1,1,0.55}
\definecolor{redy}{cmyk}{0,1,1,0.7}
\definecolor{greeny}{cmyk}{0.99,0,0.59,0.98}
\definecolor{green-go}{cmyk}{0.79,0,0.59,0.5}

\usepackage{titlesec}

\def\hhref#1{\href{http://arxiv.org/abs/#1}{arXiv:#1}} % in bibliography

 \def\Lag{\mathscr{L}}
 
\newcommand{\tmtextbf}[1]{{\bfseries{#1}}}
\newcommand{\tmtextrm}[1]{{\rmfamily{#1}}}
\newcommand{\bp}{\bar M_P}

\def\be{\begin{equation}}
\def\ee{\end{equation}}
\def\ba{\begin{array} }
\def\bac{\begin{array} {c}}
\def\bacc{\begin{array} {cc}}
\def\baccc{\begin{array} {ccc}}
\def\bacccc{\begin{array} {cccc}}
\def\ea{\end{array}}
\def\bea{\begin{eqnarray}}
\def\eea{\end{eqnarray}}

\definecolor{red}{rgb}{1,0,0}

\def\psl{\hbox{\hbox{${p}$}}\kern-1.9mm{\hbox{${/}$}}}
\def\dsl{\hbox{\hbox{${\partial}$}}\kern-2.2mm{\hbox{${/}$}}}
\def\Dsl{\hbox{\hbox{${D}$}}\kern-2.6mm{\hbox{${/}$}}}

\def\Lag{\mathscr{L}}

\newcommand{\gappeq}{{\rlap{{\raise}.5ex\text{\ensuremath{>}}}{{\lower}.5ex\text{\ensuremath{\sim}}}}}
\newcommand{\lappeq}{{\rlap{{\raise}.5ex\text{\ensuremath{<}}}{{\lower}.5ex\text{\ensuremath{\sim}}}}}

\newcommand{\I}{\tmtextrm{1{\kern}-.24em l}}

\begin{document}
\topmargin -1.0cm
\oddsidemargin -0.5cm
\evensidemargin -0.5cm

{\vspace{-1cm}}
\begin{center}

\vspace{-1cm}

 \begin{center}CERN-TH-2017-068\end{center}
 
\vspace{1cm}

 {\Huge \tmtextbf{ 
\color{blus}Inflationary Perturbations   in No-Scale Theories}} {\vspace{.5cm}}\\
 
\vspace{1.1cm}

{\large  {\bf Alberto Salvio }
%\vspace{.4cm}\\
%{\large }\\
\vspace{.3cm}
{\em  

CERN, Theoretical Physics Department, Geneva, Switzerland\\

\vspace{0.2cm}

 \vspace{0.5cm}
}

%to do
%REad it all
%check spelling 
%estetica (allineamento  delle formule, etc )

\vspace{0.5cm}

}
\vspace{0.1cm}
 \end{center}
\noindent --------------------------------------------------------------------------------------------------------------------------------

% \begin{abstract}
\begin{center}
{\bf \large Abstract}\\ 
\end{center}
{\large  We study the inflationary perturbations in general (classically) scale-invariant theories. Such scenario is motivated by the hierarchy problem and provides natural inflationary potentials and dark matter candidates.  We analyse in detail all sectors (the scalar, vector and tensor perturbations) giving general formul$\ae$ for the potentially observable power spectra, as well as for the curvature spectral index $n_s$ and the tensor-to-scalar ratio $r$.  We show that the conserved Hamiltonian for all perturbations does not feature negative energies even in the presence of the Weyl-squared term if the appropriate quantization is performed and argue that this term does not lead to phenomenological problems at least in some relevant setups. The general formul$\ae$ are then applied to a concrete no-scale model, which includes  the higgs  and a scalar, ``the planckion", whose vacuum expectation value generates the Planck mass. Inflation can be triggered by a combination of the planckion and the Starobinsky scalar and we show that no tension with observations is present even in the case of pure planckion-inflation, if the coefficient of the Weyl-squared term is large enough. In general, even quadratic inflation is allowed in this case. Moreover, the Weyl-squared term leads to an isocurvature mode, which currently satisfies the observational bounds, but may be detectable with future experiments.  }

\noindent \normalsize

%\end{abstract}

\vspace{.5cm}

\noindent --------------------------------------------------------------------------------------------------------------------------------

\vspace{1cm}

\noindent Email: alberto.salvio@cern.ch 

\newpage

 \tableofcontents

\newpage
\section{Introduction}

Theories with (classical) scale-invariance provide a dynamical origin of all mass scales \cite{Adler:1982ri,Coleman:1973jx,Salvio:2014soa,Einhorn:2014gfa,Einhorn:2016mws,Einhorn:2015lzy} and present a  number of interesting aspects. They lead to naturally flat inflationary potentials \cite{Khoze:2013uia,Kannike:2014mia,Rinaldi:2014gha,Salvio:2014soa,Kannike:2015apa,Kannike:2015fom,Barrie:2016rnv,Tambalo:2016eqr} and dark matter candidates \cite{Hambye:2013sna,Karam:2015jta,Kannike:2015apa,Kannike:2016bny,Karam:2016rsz} and represent an interesting framework to address the hierarchy problem \cite{Bardeen,Foot:2007iy,AlexanderNunneley:2010nw,Englert:2013gz,Hambye:2013sna,Farzinnia:2013pga,Altmannshofer:2014vra,Holthausen:2013ota,Salvio:2014soa,Einhorn:2014gfa,Kannike:2015apa,Farzinnia:2015fka,Kannike:2016bny}. This no-scale principle has also the virtue of being predictive as only dimension-four operators are allowed in the classical Lagrangian, which can be viewed  as a strong constraint on the allowed free parameters.

The absence of mass parameters implies that the pure gravitational piece of the Lagrangian is quadratic in the curvature tensors. The most general gravitational Lagrangian can be shown to be the sum of the Ricci scalar squared $R^2$ and the Weyl-squared terms  (modulo total derivatives). In order to solve the hierarchy problem in this framework including gravity  it is necessary that the coefficients of these two terms be large enough, which corresponds to small enough gravitational couplings \cite{Salvio:2014soa,Kannike:2015apa}. The lower bound on the coefficient of $R^2$  depends on the non-minimal coupling between $R$ and the scalar fields (another possible scale-invariant term in the Lagrangian) \cite{Kannike:2015apa}. The lower bound on the Weyl-squared term is instead model-independent and, as we will see, leads to potentially observable effects in cosmology.

Therefore, if one supposes that nature does not have fundamental scales and all observed masses are in fact dynamically generated one is   led to conjecture that the fundamental  theory of gravity  contains only terms quadratic in curvature in addition to all possible scale-invariant couplings to matter. Moreover, the requirement that this  four-derivative gravity theory cures the Higgs naturalness problem leads to the possibility of  testing this hypothesis with observations of the early universe. 

This scenario is indeed known to provide a  renormalizable theory of gravity and of all other interactions \cite{Stelle:1977ry,Salvio:2014soa}. The price to pay is the  presence of a ghost, a field whose quanta should include negative norms. This field has spin-2 and a mass $M_2$ fixed by the coefficient of the Weyl-squared term; the above-mentioned naturalness bound corresponds to $M_2\lesssim 10^{11}$ GeV \cite{Salvio:2014soa}, realising in an explicit theory the softened gravity idea of \cite{Giudice:2014tma}: for energies below $M_2$ one recovers general relativity coupled to an ordinary matter sector \cite{Salles:2014rua}, while above  the threshold $M_2$ gravity gets softened compared to the behaviour in  Einstein's gravity. It has been proposed that such theory could make sense above $M_2$ as a Lee-Wick theory \cite{Lee:1969fy,Hasslacher:1980hd,Avramidi:1985ki,Antoniadis}: the ghost is unstable and does not appear among the asymptotic states, leading to a unitary S-matrix. Also, there has been recent progress on the quantum mechanics of four-derivative theories \cite{Salvio:2015gsi,Raidal:2016wop} and a renewed further  interest in four-derivative gravity  \cite{Lu:2015cqa,Lu:2015psa,Mauro:2015waa,Alvarez-Gaume:2015rwa,Maggiore:2015rma,Barvinsky:2015uxa,Holdom:2015kbf,Salvio:2016vxi,Salvio:2016mvj,Donoghue:2016xnh,Holdom:2016nek,Narain:2011gs,Narain:2016sgk}.
Refs. \cite{Coleman,Grinstein:2008bg} claimed, however, that the Lee-Wick option might result in a violation of causality. Our approach to this problem in the present work is practical: we wish to understand whether such particle can be compatible with the available observations of the early universe we have. We leave the analysis of the remaining theoretical issues for future work.

The present work is intended to be a complete study of inflationary perturbations in general scale-invariant theories (which will be defined in detail in section \ref{Scale invariant theories}). Indeed, the presence of a naturally flat potential is only a good starting point for successful inflation. The observational quantities that can be measured are the result of tiny quantum perturbations generated during inflation, which have been later amplified and whose effects are observable today (see \cite{Mukhanov:2005sc} and \cite{Weinberg:2008zzc} for a text-book introduction).  Partial results on this subject have already been published  \cite{Deruelle:2010kf,Deruelle:2012xv,Myung:2014jha,Myung:2014cra,Myung:2015vya,Kannike:2015apa,Clunan:2009er,Ivanov:2016hcm,Tokareva:2016ied}, but the present work covers the general  set of classically scale-invariant theories. Indeed, our analysis includes a large class of models such as the most general two-derivative theories with an arbitrary number of scalar fields  and a detailed discussion of the role of the Weyl-squared term. The bound from a natural electroweak (EW) scale, $M_2\lesssim 10^{11}$ GeV, suggests that such term becomes relevant whenever the Hubble rate during inflation exceeds roughly the scale of $10^{11}$ GeV.
 The results we find actually also hold  for some non-scale-invariant theories, they apply to generic renormalizable models. Indeed,  after scale invariance is broken dynamically all the scales, such as the Planck scale, the EW scale and  the cosmological constant, appear. Given that we include them as effective mass parameters in the Lagrangian we are able to show that the results we find hold for scale-invariant as well as general renormalizable models.  We also revisit the studies that already appear in the literature; in some cases we confirm the previous results with improved derivations, while in other cases we correct some expressions for the perturbations.

After a general discussion of the perturbations in section \ref{Perturbations (generalities)}, the scalar, vector and tensor perturbations are all analysed in detail (in sections \ref{Scalar perturbations}, \ref{Vector perturbations},  \ref{Tensor perturbations}, respectively) and the relevant degrees of freedom are identified.   We explain how the known perturbations in Einstein's gravity coupled to matter are reproduced when $M_2$ is much bigger than  the Hubble rate  during inflation.  We use both a Lagrangian and a Hamiltonian approach (which are introduced in appendix \ref{Lagrange and Hamilton} for four-derivative theories). The Lagrangian one is used to derive the form of the perturbations, while the Hamiltonian formalism (developed in appendix \ref{Hamiltonian approach}) allows us to show that the full conserved Hamiltonian of the perturbations  does not feature negative energies if appropriately quantized. 

Among the most important results we find there are  the expressions (given in section \ref{Observational quantities})  for the potentially observable quantities in a general no-scale theory: the curvature and  tensor  power spectra  (with the derived formul$\ae$ for the tensor-to-scalar ratio $r$ and the curvature spectral index $n_s$) as well as a detailed discussion of the power spectra of the other perturbations. 
In particular the Weyl-squared term can lead (depending on the size of its coefficient) to an isocurvature mode whose amplitude is typically small as it turns out to be smaller than the tensor amplitude in Einstein's theory.
% We rederive the slow-roll conditions and present the scalar spectral index $n_s$ and the tensor-to-scalar ratio $r$ in Einstein gravity coupled to a generic number of scalar fields with non-trivial field metric. We extend the calculation to include the Weyl-squared term in the Lagrangian giving formul$\ae$ for $n_s$ and $r$ that are valid in the most general (classically) scale invariant theory. 

In section \ref{An example: the Higgs and the Higgs of gravity} we also apply these results to a specific model with all terms quadratic in curvature, the Higgs field and a scalar that generates the Planck scale (and therefore dubbed  ``the planckion"). This model can satisfy the most recent  observational bounds of Planck \cite{Ade:2015lrj} on $n_s$ and $r$ as well as on the isocurvature power spectra. Interestingly, when the coefficient of the Weyl-squared term is large enough  even an inflation due to a quadratic potential is allowed; for example, an inflation triggered by the planckion is permitted, eliminating a tension that was previously found neglecting the Weyl-squared term \cite{Kannike:2015apa}. In this case the  predictions of  planckion-inflation are close to Planck's observational bounds on isocurvature perturbations, which suggests that this possibility can be tested with future observations. 

Finally, in section \ref{Issues due to the Weyl term}, we argue that the possible issues due to the ghost, which we have discussed above, do  not lead to phenomenological problems in some no-scale models (including the one we have just mentioned), at least if one saturates the bound on $M_2$ required by the solution of the hierarchy problem.
%UP TO HERE

\section{Scale-invariant theories} \label{Scale invariant theories}

In this work we consider general (classically) scale-invariant theories. The action is 
\be S=\int d^4x  \sqrt{|\det g|} \Lag, \qquad \Lag=  \mathscr{L}_{\rm gravity} + \mathscr{L}_{\rm matter}+  \mathscr{L}_{\rm non-minimal} . \ee

The term $ \mathscr{L}_{\rm gravity}$ is the pure gravitational piece. The no-scale principle dictates that it is given by the most general Lagrangian quadratic in curvature,
 \be  \mathscr{L}_{\rm gravity} = \alpha R^2  +\beta R_{\mu\nu}^2 +\gamma R_{\mu\nu\rho\sigma}^2.
 \ee 
 The dimensionless parameters $\alpha$, $\beta$ and $\gamma$ are not all independent because of the Gauss-Bonnet relation in four-dimensions according to which \cite{Stelle:1977ry}
 \be  \sqrt{|\det g|}\left(R^2  -4 R_{\mu\nu}^2 +  R_{\mu\nu\rho\sigma}^2\right) = \mbox{divergence}. \ee 
Therefore, we can restrict our attention to $R^2$ and $R_{\mu\nu}^2$ and write their most general linear combination as  
\be \mathscr{L}_{\rm gravity} =   \frac{\frac13 R^2-R_{\mu\nu}^2}{f_2^2} +\frac{R^2}{6f_0^2}.  \ee
We have grouped together $\frac13 R^2-R_{\mu\nu}^2$ because this quantity is (up to total derivatives) proportional to the squared of the Weyl tensor $W_{\mu\nu\rho\sigma}$:
\be \int d^4x  \sqrt{|\det g|} \,\bigg[\frac{\frac13 R^2-R_{\mu\nu}^2}{f_2^2}  \bigg] = -\frac12  \int d^4x  \sqrt{|\det g|} W_{\mu\nu\rho\sigma}^2.  \ee

The piece  $\mathscr{L}_{\rm non-minimal}$ represents the non-minimal couplings between the scalar fields $\phi^a$  and the Ricci scalar 
\be \mathscr{L}_{\rm non-minimal} = -\frac12 \xi_{ab} \phi^a\phi^b R.\ee
Finally,  $\mathscr{L}_{\rm matter}$ is the remaining part of the Lagrangian, which depends on the matter fields: the gauge fields $A^B_\mu$ (with field strength $F_{\mu\nu}^B$),  the fermions $\psi_j$ and the  scalars $\phi^a$. The absence of dimensionful parameter is so restrictive that we can write their Lagrangian in one line:
 \be \mathscr{L}_{\rm matter} =  - \frac14 (F_{\mu\nu}^B)^2 + \frac{(D_\mu \phi^a)^2}{2}  + \bar\psi_j i\slashed{D} \psi_j -
(Y^a_{ij} \psi_i\psi_j \phi^a + \hbox{h.c.}) - \frac{\lambda_{abcd}}{4!} \phi^a\phi^b\phi^c\phi^d. \label{Lmatter}
%\frac{1}{2} \partial_\mu \phi^a \partial^\mu \phi^a -  \lambda_{abcd} \phi_a\phi_b\phi_c\phi_d/4! 
\ee

All the coefficients defined so far are dimensionless and so respect the no-scale principle. However, the scales we observe in nature, such as the Planck or the weak scale, must of course  be generated. This can occur in two alternative ways: perturbatively or non-perturbatively. In the first case we assume that some scalar field(s) acquire a vacuum expectation value (VEV) which gives rise to the  mass scales \cite{Salvio:2014soa}; in section \ref{An example: the Higgs and the Higgs of gravity} we will illustrate an example of this type. In the second way one supposes  that some strongly coupled sector (such as an SU($n$) gauge theory) confines and thus generates the observed scales through its coupling to the other sectors (e.g. its gravitational couplings)  \cite{Adler:1982ri}. After this has happened the Planck mass, the weak scale, the cosmological constant, etc appear in the Lagrangian as effective parameters. We will therefore introduce these quantities directly in the action in the following sections. This will also allow us to be more general and cover arbitrary renormalizable models.

\subsection{The action in the Einstein frame}\label{Einstein frame}

Since our task is to study inflationary perturbations, from now on we restrict our attention to the scalar-tensor sector of the theory. Of course,  fermions and gauge fields should also be present (with a Lagrangian of the form (\ref{Lmatter})) both for phenomenological reasons and to realize the dynamical generation of scales that we have discussed \cite{Adler:1982ri,Salvio:2014soa,Kannike:2015apa}.

The most general renormalizable  scalar-tensor theory is  (up to total derivatives)
\be S_{st} = \int d^4x  \sqrt{|\det g|} \,\bigg[\frac{\frac13 R^2-R_{\mu\nu}^2}{f_2^2} +\frac{R^2}{6f_0^2} -\frac{\bar M_{\rm Pl}^2+F(\phi)}{2}R+ \frac{1}{2} \left(\partial_\mu \phi^a\right)^2 - V(\phi) \bigg],\label{renAction}\ee
where $\bp$ is the reduced Planck mass.
Renormalizability requires $F(\phi) =  \xi_{ab} \phi^a\phi^b+...$ 
%$Z_{ab}(\phi) = \delta_{ab}$ 
and $V(\phi) =  \lambda_{abcd} \phi^a\phi^b\phi^c\phi^d/4!+...$ (where the dots are terms with lower powers of $\phi^a$) in the bare Lagrangian; however, we will keep these functions general in the following  to take into account possible field dependence of the couplings induced by the renormalization group. As we have discussed in the introduction,  the presence of the terms quadratic in curvature promotes gravity to a renormalizable theory, but also introduces a spin-2 ghost. The mass of this field is  $M_2 = f_2\bp/\sqrt{2}$ \cite{Stelle:1977ry}. One of the motivations of this work is to understand whether the presence of this ghost is consistent with the observations of the early universe  we have available.

To proceed further it is useful to rewrite the action in a more familiar form. We follow here the method presented in \cite{Kannike:2015apa} appropriately extended to take into account the effective massive parameters. The non-standard $R^2$ term can be removed by introducing an auxiliary field $\chi$ with a Lagrangian that vanishes on-shell:
\be  - \sqrt{|\det g|} \frac{(R+3f_0^2 \chi/2)^2}{6f_0^2},\ee
which we are therefore free to add to the total Lagrangian. Once we have done so   the action reads 
\be S_{st} = \int d^4x  \sqrt{|\det g|} \,\bigg[ \frac{\frac13 R^2-R_{\mu\nu}^2}{f_2^2} -\frac{f(\chi,\phi)}{2} R + \frac{1}{2}\left(\partial_\mu \phi^a\right)^2 - V(\phi) -\frac{3f_0^2\chi^2}{8} \bigg],\ee
where $f(\chi,\phi)\equiv\bp^2 + F(\phi)+\chi$.
In order to get rid of the remaining non-standard $fR$ term we perform a conformal transformation of the metric 
\be g_{\mu\nu} \rightarrow \frac{\bp^2}{f}g_{\mu\nu} \label{ConfTransf}\ee
and the action becomes 
\be S_{st} = \int d^4x  \sqrt{|\det g|} \,\left\{ \frac{\frac13 R^2-R_{\mu\nu}^2}{f_2^2}   -\frac{\bar M_{\rm Pl}^2}{2} R  +\bp^2 \left[\frac{\left(\partial_\mu \phi^a\right)^2}{2f} +\frac{3(\partial_\mu f)^2}{4f^2}\right]- U \right\}, \label{SstTransf}\ee
where 
\be U = \frac{\bp^4}{f^2}\left( V +\frac{3f_0^2}{8} \chi^2\right). \ee
The form of the action in (\ref{SstTransf}) is known as the Einstein frame because all the non-minimal couplings have been removed.
The field $f$ can now be seen as an extra scalar degree of freedom. It is interesting to note that the remaining parts of the action, which depend on the fermions and the gauge fields, remain invariant under the conformal transformation if we also transform the matter fields as follows:
\be \phi^a \rightarrow \left(\frac{f}{\bp^2}\right)^{1/2} \phi^a, \qquad  \psi_j \rightarrow \left(\frac{f}{\bp^2}\right)^{3/4}  \psi_j, \qquad A_\mu^B \rightarrow  A_\mu^B. \ee
However, the scalar kinetic terms are not invariant, so we have not performed this transformation  in (\ref{SstTransf}).
 To simplify further the action we define $\zeta=\sqrt{6f}$ (notice that in order for the metric redefinition in (\ref{ConfTransf}) to be regular one has to have $f>0$ and thus we can safely take the square root of $f$) and we obtain 
\be S_{st} = \int d^4x  \sqrt{|\det g|} \,\left\{ \frac{\frac13 R^2-R_{\mu\nu}^2}{f_2^2}   -\frac{\bar M_{\rm Pl}^2}{2} R  +\frac{6\bp^2}{2\zeta^2} \left[\left(\partial_\mu \phi^a\right)^2  + (\partial_\mu \zeta)^2 \right]- U(\zeta,\phi) \right\},\ee
where 
\be U(\zeta,\phi) = \frac{36 \bp^4}{\zeta^4}\left[V(\phi) +\frac{3f_0^2}{8} \left(\frac{\zeta^2}{6} - \bp^2 - F(\phi)\right)^2\right]. \ee

Therefore, we have been able to write the action as the sum of two pieces:
\be S= S_W + S_{ES}, \ee
where $S_W$ is  the part due to the unusual Weyl-squared term,
\be S_W=\int d^4x  \sqrt{|\det g|} \,\bigg[\frac{\frac13 R^2-R_{\mu\nu}^2}{f_2^2} \bigg], \label{action2}
 \ee
 and $S_{ES}$ is the $E$instein-Hilbert part plus the $S$calar field piece equipped with a non-trivial field metric,
\be S_{ES}=\int d^4x  \sqrt{|\det g|} \,\bigg[ -\frac{\bar M_{\rm Pl}^2}{2}R  + 
\frac{K_{ij}(\phi) }{2}\partial_\mu \phi^i\partial^\mu \phi^j-
U(\phi)
\bigg]. 
 \ee
 Here, for notational simplicity,  we have introduced a new set of fields $\phi^i$ where the index $i$ runs over the possible values of the index $a$ plus $\zeta$, in other words $\phi^i=\{\phi^a, \zeta\}$. Also the field metric is given by 
 \be K_{ij} =  \frac{6\bp^2}{\zeta^2} \delta_{ij}. \label{KijRen}\ee
 
 \subsection{FRW background and slow-roll inflation}
 
In this section we consider a Friedmann-Robertson-Walker (FRW) metric
 \be ds^2 = dt^2 -a(t)^2 \left[dr^2+r^2(d\theta^2 +\sin^2\theta d\phi^2)\right],  \ee
 where $a$ is the scale factor
and we have neglected  the spatial curvature parameter as during inflation the energy density is dominated by the scalar fields. The hypothesis of homogeneity and isotropy will be relaxed in the next section where the perturbations around the FRW metric will be considered.

For the following analysis of the perturbations  it is convenient to introduce the conformal time $\eta$ defined as usual  by
\be \eta = \int_{t^*}^t \frac{dt'}{a(t')}, \ee
where $t^*$ is some reference time; in the following we will choose $t^* \rightarrow \infty$. The FRW metric in terms of $\eta$  is
  \be ds^2 = a(\eta)^2\left(d\eta^2 -\delta_{ij}dx^idx^j\right).  \ee
  
 The Einstein equations  are   \bea {\cal H}^2&=&\frac{K_{ij} \phi'^i \phi'^j/2+a^2U}{3 \bar M_{\rm Pl}^2} ,\label{EOM1c} \\ 
{\cal H}^2-{\cal H}'&=& \frac{K_{ij} \phi'^i \phi'^j}{ 2\bar M_{\rm Pl}^2}, \label{EOM2c}\eea
 where ${\cal H} \equiv a'/a$ (related to $H\equiv \dot a/a$ by ${\cal H} = aH$) and a prime denotes a derivative with respect to $\eta$, while a dot is a derivative with respect to $t$. The equations for the scalar fields are instead
 \be \phi''^i +\gamma^i_{jk} \phi'^j \phi'^k +2{\cal H} \phi'^i+a^2U^{,i}=0. \ee
 Here  for a generic function $F$ of the scalar fields, we defined $F_{,i}\equiv \partial F/\partial \phi^i$, also $\gamma^i_{jk}$ is the affine connection in the scalar field space
 \be \gamma^i_{jk}\equiv \frac{K^{il}}{2}\left(K_{lj,k}+K_{lk,j}-K_{jk,l}\right) \ee
 and $K^{ij}$ denotes the inverse of the field metric (which is used to raise and lower the scalar indices $i,j,k, ...$); for example $F^{,i}\equiv K^{ij}F_{,j}$. 
 
 Notice that in the case of pure de Sitter space-time ($a(\eta)=-1/(H\eta)$, $\phi'^i=0,$ $U_{,i} =0$) eq. (\ref{EOM1c}) tells us 
  $ {\cal H}^2 =  a^2 U/(3\bp^2)$ and Eq. (\ref{EOM2c}) implies 
  ${\cal H}' = {\cal H}^2$.

   In the slow-roll regime we require the scalar equation to reduce to 
   \be   \phi'^i\approx -\frac{a^2}{3{\cal H}}U^{,i} \label{ScFiEq}\ee
   so 
   \be \phi''^i +\gamma^i_{jk} \phi'^j \phi'^k\approx {\cal H} \phi'^i. \ee
Moreover, 
 \be {\cal H}^2\approx\frac{a^2U}{3 \bar M_{\rm Pl}^2}. \label{Hslow-roll}\ee
The slow-roll occurs when two conditions are satisfied \cite{Salvio:2016vxi} (see also \cite{Chiba:2008rp} for previous studies):
\be \epsilon \equiv  \frac{  \bar M_{\rm Pl}^2 U_{,i}U^{,i}}{2U^2} \ll 1. \label{1st-slow-roll}\ee 
\be \left|\frac{\eta^{i}_{\,\,\, j} U^{,j}}{U^{,i}}\right|  \ll 1 \quad \mbox{($i$  not summed), }\quad \mbox{where}\quad \eta^{i}_{\,\,\, j}\equiv \frac{\bar M_{\rm Pl}^2 U^{;i}_{\,\,\, ;j}}{U}. \label{2nd-slow-roll} \ee
 It is easy to check that $\epsilon$ and $\eta^i_{\,\,\, j}$ reduce to the well-known single field slow-roll parameters in the presence of only one field.

We now introduce the number of e-folds for a generic multi-field theory. By writing the equations in (\ref{ScFiEq}) and (\ref{Hslow-roll}) in terms of the cosmic time $t$ we obtain the following dynamical system for $\phi^i$: 
\be \phi^i=-\frac{\bar M_{\rm Pl}U^{,i}(\phi)}{\sqrt{3U(\phi)}}, \label{dynamical1}\ee
which can be solved with a condition at some initial time  $t_0$: namely $\phi^i(t_0)=\phi^i_0$. Once the functions $\phi^i(t)$ are known we can obtain $H(t)$ from eq. (\ref{Hslow-roll}).
The number of e-folds $N$ can be introduced by 
\be N(\phi_0) \equiv \int_{t_e}^{t_0(\phi_0)} dt' H(t'), \label{Ndef}\ee
where $t_e$ is the time when inflation ends. Dropping the label on $t_0$ and $\phi_0$ as they are generic values we have
\be N(\phi) \equiv \int_{t_e}^{t(\phi)} dt' H(t'). \label{Ndef2}\ee
Notice that we write $t$ as a function of $\phi$: this is because once the initial position $\phi$ in field space is fixed the time required to go from $\phi$ to the field value when inflation ends is fixed too because the dynamical system in (\ref{dynamical1}) is of the first order. Note, however, that $H$ also generically depends on $\phi$.
% Definition (\ref{Ndef2}) implies
%\be \frac{dN}{dt}=H,\label{dN/dt}\ee which can be used in (\ref{slow-roll-eq}) to obtain a simpler dynamical system for $\phi^i$ where the independent variable is $N$ instead of $t$: 
%\be \frac{d\phi^i}{dN}=- \frac{\bar M_{\rm Pl}^2U^{,i}(\phi)}{U(\phi)}.\ee 

 \section{Perturbations (generalities)}\label{Perturbations (generalities)}

By  choosing the conformal Newtonian gauge, the metric describing the small fluctuations around the FRW space-time can be written as 
\be ds^2= a(\eta)^2\left\{ (1+2\Phi(\eta, \vec{x})) d\eta^2 -2 V_i(\eta, \vec{x}) d\eta dx^i  - \left[(1-2\Psi(\eta, \vec{x})) \delta_{ij}+h_{ij}(\eta, \vec{x})\right]dx^idx^j\right\}. \label{dsPert}\ee
Where, by definition,  the vector perturbations $V_i$ satisfy
\be \partial_iV_i=0 \label{Cond1}\ee
 and the tensor perturbations obey
 \be h_{ij}=h_{ji},  \qquad h_{ii} =0, \qquad \partial_ih_{ij}=0. \label{Cond2}\ee
%\be g_{\eta\eta} =a^2( 1+2\Phi), \quad g_{\eta i}=0, \quad g_{ij} = - a^2 \delta_{ij} (1-2\Psi(\eta, \vec{x})).   \ee 
Sometimes the Newtonian gauge is defined for the scalar perturbations $\Phi$ and $\Psi$ only  (see e.g. \cite{Weinberg:2008zzc}). Here we considered an extended definition, which also includes the non-scalar perturbations. A possible gauge dependent divergence of $h_{ij}$ has been set to zero by appropriately choosing the gauge.

Also we decompose the scalar fields $\phi^i(\eta, \vec{x})$  in the background  $ \phi^i(\eta)$, which are only time-dependent, plus the fluctuations $\varphi^i(\eta, \vec{x})$,
\be \phi^i(\eta, \vec{x}) \rightarrow \phi^i(\eta) + \varphi^i(\eta, \vec{x}) . \ee 

As it is well-known there are no mixing between tensor, vector and scalar perturbations from the   part $S_{ES}$ of the action. The same is  also true for the Weyl term $S_W$. Indeed, that property only follows from (\ref{Cond1}) and (\ref{Cond2}) and rotation invariance. We therefore study the various sectors separately in the following. 

Previous studies of the perturbations  in less general setups can be found in \cite{Deruelle:2010kf,Deruelle:2012xv,Myung:2014jha,Myung:2014cra,Myung:2015vya,Clunan:2009er,Ivanov:2016hcm,Tokareva:2016ied}. We will also revisit these studies  and find some differences with some of them, which will be discussed in the following.

 \section{Scalar perturbations}\label{Scalar perturbations}

Let us start with the scalar perturbations, whose quadratic action we denote with $S^{(S)}$. This action has a contribution from the Weyl-squared term and one from the remaining terms,  $S^{(S)}=S^{(S)}_W+S^{(S)}_{ES}$, where 
\bea S^{(S)}_W &=& -\frac{2}{3f_2^2}\int d^4 x \left[\vec{\nabla}^2\left(\Phi+\Psi\right) \right]^2, \label{SW} \\ 
S^{(S)}_{ES} &=&\int d^4 x\frac{a^2}{2}  \left\{\bp^2\left[-6 \Psi'^2 -12 {\cal H} \Phi \Psi' +4 \Psi \vec{\nabla}^2 \Phi -2 \Psi \vec{\nabla}^2 \Psi - 2 \left( {\cal H}'+2{\cal H}^2 \right) \Phi^2 \right]\right. \nonumber  \\ 
&&+K_{ij} \left(\varphi'^i\varphi'^j +\varphi^i\vec{\nabla}^2\varphi^j \right) +2 K_{ij,l} \phi'^i\varphi'^j \varphi^l -(\Phi+3\Psi) \left(2K_{ij} \phi'^i \varphi'^j + K_{ij,l} \phi'^i\phi'^j\varphi^l\right) \nonumber \\ 
&& \left.-a^2 U_{,i,j}\varphi^i\varphi^j -2a^2(\Phi-3\Psi) U_{,i} \varphi^i \right\}.  \label{SEH}  \eea
 In order to derive $S^{(S)}_{ES}$ we have used the background equations (\ref{EOM1c}) and (\ref{EOM2c}) and we have dropped total derivatives. Notice that the field metric $K_{ij}$ is totally general in this expression and does not have to satisfy eq. (\ref{KijRen}), which is characteristic of renormalizable theories. 
   
  The fact that the time-derivative of the perturbation $\Phi$ does not appear in the action above tells us that it should be considered as a non-dynamical field.
  
  One might wonder why there are no more than two time-derivatives in the action for scalar perturbations even if we started from an action with four derivatives. The reason is that in Einstein gravity there is no independent degrees of freedom in the scalar sector, while with the addition of the Weyl-squared term we should find one degree of freedom (the one corresponding to the helicity-0 component of the massive spin-2 field). If we had found four time-derivatives in the scalar action  we would instead have  two scalar degrees of freedom instead of one because a four-derivative system can always been interpreted as a two-derivative system with the number of degrees of freedom doubled.

  \subsection{Pure de Sitter} \label{Pure de Sitter} 
  
The expression in (\ref{SEH}) simplifies considerably in the case of pure de Sitter space-time, $a(\eta)=-1/(H\eta)$, for which ${\cal H}' = {\cal H}^2$, $\phi'^i=0$ and  $U_{,i} =0$. Moreover, we know that the de Sitter space-time is a reasonably good approximation during inflation because we assume that the slow-roll conditions in (\ref{1st-slow-roll}) and (\ref{2nd-slow-roll}) are satisfied. We therefore consider this case in the present section. In the next one we will study the small departures from this space-time due to the small, but non-zero time-dependence of the scalar fields. 
  
  \subsubsection{Scalar field perturbations}

  We start from the scalar field perturbations $\varphi^i$. We now show that, thanks to the special expression of the field metric in (\ref{KijRen}) that is realized for {\it any} renormalizable model, the mixing between the different  $\varphi^i$ can be eliminated with a field redefinition and one can analize the various  $\varphi^i$ separately. 
  
  The field equations of  $\varphi^i$ are
  \be K_{ij} \varphi''^j + 2{\cal H} K_{ij} \varphi'^j - K_{ij}\vec{\nabla}^2\varphi^j + a^2 U_{,i,j} \varphi^j = 0.\ee
%  where we have expanded the fields on spatial plane waves $\varphi^i \propto  e^{i \vec{k} \cdot \vec{x}}$. 
We now multiply this equation by $K^{il}$ and sum over $i$, to obtain
  \be  \varphi''^l+ 2{\cal H}  \varphi'^l - \vec{\nabla}^2 \varphi'^l +U_{,i,j} K^{il} \varphi^j = 0.   \ee
  Notice that the matrix $m^2$ with elements $m^{2\,l}_{\,\,\, \,\, j} \equiv  U_{,i,j} K^{il}$ is symmetric as a consequence of the symmetry of $U$, i.e.  $U_{,i,j} = U_{,j,i}$ and the proportionality between $K_{ij}$ and $\delta_{ij}$ (equation (\ref{KijRen})). Therefore we can always diagonalize $m^2$ with an orthogonal transformation, $\varphi \rightarrow {\cal O} \varphi$, and, after this transformation, the equation for $\varphi$ is (suppressing the index $i$ for simplicity)
  \be \varphi''+2{\cal H} \varphi'+(m^2 a^2-\vec{\nabla}^2) \varphi = 0 \label{EOMvarphi} \ee
   and the corresponding action is (rescaling the field to have a canonically normalized kinetic term)
\be S_{\varphi}=  \int d^4 x \Lag_{\varphi}, \qquad \mbox{where}\qquad   \Lag_{\varphi} = \frac{a^2}{2}  \left\{\varphi'^2 + \varphi\vec{\nabla}^2\varphi  -m^2 a^2\varphi^2 \right\}. \label{LagVarphi}\ee
%Now $m^2$ represent one of the eigenvalues of the mass matrix $m^2$. 

We can now quantize the theory with the standard canonical procedure. Of course, it is well-known how to do this for a scalar field on de Sitter space. Nevertheless, here we revisit such procedure in a way that will be useful to study %We review this procedure now  and later on we will extend it to 
 $\Psi$ as well as the vector and tensor perturbations, which, as we will see, requires an unusual quantization in the presence of the Weyl-squared term. We   introduce the conjugate momentum 
\be \pi_{\varphi} = \frac{\partial\Lag_{\varphi}}{\partial \varphi'} = a^2 \varphi' \label{ConjVarphi}\ee
and we impose the canonical quantization conditions: 
\be [\varphi(\eta,\vec{x}), \varphi'(\eta,\vec{y})] = \frac{i}{a^2} \delta^{(3)}(\vec{x} - \vec{y}), \quad [\varphi(\eta,\vec{x}), \varphi(\eta,\vec{y})] = 0, \quad [\varphi'(\eta,\vec{x}), \varphi'(\eta,\vec{y})] = 0.  \label{canComm} \ee

We expand now the field by considering the Fourier transform with respect to $\vec{x}$, but not on $\eta$, that is
\bea \varphi(\eta,\vec{x}) =  \int \frac{d^3q}{(2\pi)^{3/2}}  e^{i\vec{q}\cdot \vec{x}} \varphi_0(\eta,\vec{q}).  \label{ExpMomphi}  \eea
 The hermiticity condition on $\varphi(\eta,\vec{x})$, i.e.  $\varphi(\eta,\vec{x})^\dagger=\varphi(\eta,\vec{x})$, implies 
  \be  \varphi_0(\eta, \vec{q})^\dagger =   \varphi_0(\eta, -\vec{q}) \label{HermMode}\ee
%\be \varphi(\eta,\vec{x}) =  \int \frac{d^3q}{(2\pi)^{3/2}}a_{\vec{q}} \varphi_{\vec{q}}(\eta) e^{i \vec{q}\cdot\vec{x}} + \mbox{h.c.}  \ee
and the equation of motion in (\ref{EOMvarphi}) dictates that  $\varphi_0$ satisfies  %$\tilde f_0(\eta,\vec{q}) a (\vec{q})$ 
%where $\varphi_{\vec{k}}(\eta)$ 
  \be  \varphi''_{0}+2{\cal H} \varphi'_{0}+(m^2 a^2+q^2) \varphi_{0} = 0. \label{phik-eq} \ee
  The general solution of (\ref{phik-eq}) is a linear combination of 
  \be \eta ^{3/2} J_{\frac{1}{2} \sqrt{9-\frac{4 m^2}{H^2}}}(q \eta) \qquad \mbox{and} \qquad \eta ^{3/2} Y_{\frac{1}{2} \sqrt{9-\frac{4 m^2}{H^2}}}(q \eta ), \label{massiveModes} \ee
  where $J_ n(z)$  and $Y_n(z)$ are the  the Bessel function of the first and second kind, respectively. 
  Since  in the superhorizon limit, $\eta \rightarrow 0^-$, we have that  the de Sitter scale factor $a^2=1/(H \eta)^2$ diverges we see that the massive fields, $m\neq 0$, are suppressed in that limit and therefore we can focus on the effectively massless degrees of freedom.
  By doing so, we can take the two linearly independent solutions to be  
  \be  y_0(\eta, q)  \equiv \frac{H |\eta|}{\sqrt{2q}}\left(1-\frac{i}{q\eta}\right)e^{-iq\eta} \qquad   %\tilde f_{0}(\eta, q)^* \equiv \frac{H |\eta|}{\sqrt{2q}}\left(1+\frac{i}{q\eta}\right)e^{iq\eta}. 
  \mbox{and its complex conjugate}. \label{ModeVarphi}\ee
  So, by using the hermiticity condition in (\ref{HermMode}), we have 
  \be  \varphi_0(\eta, \vec{q}) = a_0(\vec{q}) y_0 (\eta, q) +a_0(-\vec{q})^\dagger y_0 (\eta, q)^*,  \label{fieldITOaa}  \ee
  where $a_0(\vec{q})$ is an operator, which  will be identified with the annihilation operator of the quanta of the scalar field under study.
  
 An important feature that will be useful in analysing the other non-standard sectors is that the expansion in (\ref{ExpMomphi}) and (\ref{fieldITOaa}) can be inverted to obtain $a_0$ in terms of the field\footnote{In order to verify eq. (\ref{aqExpr})
 %   one can check it for $\eta \rightarrow -\infty$ (where the massive and massless modes coincide) and
it is useful to notice  that the quantity on the right-hand side of (\ref{aqExpr}) is independent of $\eta$ (because $y_0(\eta,\vec{q}) e^{i \vec{q}\cdot\vec{x}}$ and $\varphi(\eta, \vec{x})$ are both solutions of the equations of motion) and can therefore be evaluated in the limit $\eta \rightarrow -\infty$ where it becomes simple.
%for a solution $\varphi_{\vec{q}}$ of (\ref{phik-eq}) the following quantity is independent of $\eta$: 
  %\be a^2 \int \frac{d^3q}{(2\pi)^3}\left(\varphi_{\vec{q}} \varphi_{\vec{q}}'^* e^{i\vec{q}\cdot (\vec{x}-\vec{y})} -\varphi_{\vec{q}}' \varphi_{\vec{q}}^* e^{-i\vec{q}\cdot (\vec{x}-\vec{y})}\right). \ee} 
    }

 \be a_0(\vec{q}) = ia^2(\eta) \int \frac{d^3x}{(2\pi)^{3/2}}  e^{-i \vec{q}\cdot\vec{x}}\left( y_0(\eta,\vec{q})^* \, \stackrel{\mathclap{\normalfont\mbox{\scriptsize $\leftrightarrow$}}}{\partial_\eta}\varphi(\eta,\vec{x})\right). \label{aqExpr}\ee
 This is possible because the two solutions $y_0$ and $y_0^*$ are linearly independent.
%the normalization of $\varphi_{\vec{q}}$ has been chosen here in a way that the canonical commutators in (\ref{canComm}) are satisfied provided that the operators $a_{\vec{q}}$ fulfill
Then, by using the canonical commutators in (\ref{canComm}), one finds necessarily
  \be [a_0(\vec{k}), a_0(\vec{q})^\dagger] =  \delta(\vec{k}-\vec{q}), \qquad [a_0(\vec{k}), a_0(\vec{q})]= 0.  \ee
  If we had normalized the modes in (\ref{ModeVarphi}) differently we would have found an extra factor multiplying $\delta(\vec{k}-\vec{q})$. Therefore, what fixes the normalization constants  of the modes is the requirement that the operators $a_0$ and $a_0^\dagger$ satisfy  the standard commutator relations for annihilation and creation operators. This is, however, not quite enough to identify  $a_0$ and $a_0^\dagger$ as annihilation and creation operators, respectively. The remaining step is done in
   appendix \ref{Hamiltonian approach} where we review how the positivity of the norm of the quanta created and annihilated by  $a_0^\dagger$ and $a_0$ respectively guarantees that the Hamiltonian has a  spectrum that is bounded from below. Similar considerations will be useful to analyse the other non-standard sectors.
%   We then extend this study to other non-standard sectors in the same appendix.

%(the same is also true for the massive modes, $m\neq 0$, in which case one should use the solutions in (\ref{massiveModes})).

  \subsubsection{Metric perturbations}
  
  We now turn to the metric perturbations $\Phi$ and $\Psi$.  As we have already stated, $\Phi$ should be considered as a non-dynamical field. To find the equation that fixes its value we perform the variation of $S^{(S)}$ with respect to $\Phi$, to obtain 
  \be -\frac{4}{3 f_2^2 \bp^2 a^2} \vec{\nabla}^4 \left(\Phi+\Psi\right) -6 {\cal H} \Psi'+2\vec{\nabla}^2 \Psi -6{\cal H}^2 \Phi = 0, \label{ConstraintPhi}\ee
 %where we have expanded again the fields on spatial plane waves $\Psi, \Phi \propto  e^{i \vec{k} \cdot \vec{x}}$.
 where $\vec{\nabla}^4 = (\vec{\nabla}^2)^2$.  

   The main phenomenologically interesting regime is the superhorizon limit, $\eta\rightarrow 0$, when $a\rightarrow\infty$. We therefore focus on this case first. Then the first term in (\ref{ConstraintPhi}) is small and can be treated perturbatively in an  expansion in $1/a$. The solution of (\ref{ConstraintPhi}) at next-to-leading order in this expansion  is 
   \be \Phi = \frac{1}{3{\cal H}^2} \vec{\nabla}^2\Psi-\frac{\Psi'}{{\cal H}} + \frac{2 }{9f_2^2 \bp^2 a^2 {\cal H}^2} \vec{\nabla}^4\left(\frac{\Psi'}{{\cal H}}-\Psi\right).\ee
We now insert this constraint for  $\Phi$ in Eqs. (\ref{SW}) and (\ref{SEH}) and drop all the terms that go to zero in the $a\rightarrow \infty$ limit, to obtain 
\bea \frac{\bp^2}{2} \int d^4x a^2\left[-\frac{2}{\cal H}\left(\Psi \vec{\nabla}^2\Psi\right)'  -2\Psi\vec{\nabla}^2 \Psi +\frac{2}{3{\cal H}^2}\Psi \vec{\nabla}^4 \Psi \right. \\ \nonumber \left.  +\frac{4}{3 f_2^2\bp^2 a^2}\left(\frac{2\Psi'\vec{\nabla}^4 \Psi}{\cal H} - \frac{\Psi' \vec{\nabla}^4\Psi'}{{\cal H}^2} - \Psi \vec{\nabla}^4\Psi\right)\right].\label{actionQuadPsi} \eea
The first two terms in the expression above are apparently the leading ones but notice that an integration by parts shows\footnote{In order to see the cancellation one has to use the expression $a(\eta) = - 1/(H\eta)$ vaiid in de Sitter space-time.} that they exactly cancel each other! Therefore, both the Einstein contribution (the third term in this expression) and the Weyl contribution (the one that is divided by $f_2^2$) have the same behaviour as $a\rightarrow \infty$. By taking the variation with respect to $\Psi$ and using $\mathcal{H} = -1/\eta$ one obtains the equation
\be \frac{\Psi''}{{\cal H}^2} - 2\frac{\Psi'}{\cal H} + \frac{f_2^2\bp^2 a^2}{2{\cal H}^2} \Psi = 0. \label{EqPsisuper}  \ee
Notice that in the limit where $f_2\rightarrow \infty$ one recovers Einstein's theory result for the de Sitter space, $\Psi = 0$. The last term in this equation was neither discussed nor shown in \cite{Ivanov:2016hcm}, which performed a previous analysis of the $\Psi$-sector. We observe, however, that such term is important to recover Einstein's result. The general solution of (\ref{EqPsisuper}) is a linear combination of 
\be \eta^{(-1-\sqrt{1-4M_2^2/H^2})/2}, \qquad  \eta^{(-1+\sqrt{1-4M_2^2/H^2})/2}. \label{Psibehaviour}\ee
When $M_2 = f_2 \bp/\sqrt{2} \gg H$ very intense oscillations produced by the last term in (\ref{EqPsisuper}) make effectively $\Psi$ go to zero, but for $M_2<H$ we find a growth of $\Psi$ at superhorizon scales. At the end of section \ref{Inclusion of slow-roll} we will show that this divergence is a gauge artifact by showing that no divergences are present in another gauge (the co-moving one). At the same time, however, (\ref{Psibehaviour})  means that the Weyl term dominates even in the superhorizon limit because this divergence comes from the first two terms in (\ref{EqPsisuper}), which come in turn from the Weyl term. The reason why this happens is because the apparently leading terms coming from the Einstein-Hilbert action in Eq. (\ref{actionQuadPsi}) actually cancel each other. We conclude that the superhorizon limit is a particular case of Weyl domination (up to $M_2^2/H^2$ corrections) not of Einstein domination. This somehow differs from the classification proposed in \cite{Ivanov:2016hcm}. 
Notice that if the Weyl term dominates at superhorizon scales then it dominates at any time because a large $a$ suppresses the Weyl contribution.

Therefore, we now consider the case of Weyl domination, from where we will be able to take the superhorizon limit as we have just explained.  During Weyl domination ($f_2$ small) the constraint of $\Phi$, Eq. (\ref{ConstraintPhi}),  implies 
\be \Phi= - \Psi, \label{PsiPhiRel}\ee which we insert in (\ref{SW}) and (\ref{SEH})  to obtain the action of $\Psi$:  
\be S^{(S)}_{\Psi} = \int d^4x \Lag_{\Psi}, \qquad \mbox{where} \qquad \Lag_{\Psi} = \frac{a^2}{2}\left( - \hat\Psi'^2 - \hat\Psi \vec{\nabla}^2 \hat\Psi -4 {\cal H}^2 \hat\Psi^2\right), \ee
where $\hat\Psi \equiv \sqrt{6} \bp \Psi$.
We see that $\hat\Psi$ is a ghost; it represents the helicity-0 component of the spin-2 ghost. This type of fields require a special quantization which we will discuss. %Nevertheless, in appendix \ref{Hamiltonian approach} it is shown that the Hamiltonian, when it is conserved, has a spectrum that is bounded from below if one introduces negative norm states. We proceed then to quantize the theory. 
However, the usual canonical commutators remain unchanged  \cite{Salvio:2015gsi}. 
   The conjugate momentum of $\hat\Psi$ is 
\be \Pi_{\hat\Psi} = \frac{\partial \Lag_{\Psi}}{\partial \hat\Psi'}= -a^2 \hat\Psi'  \ee
so the canonical commutators are 
\be [\hat\Psi(\eta,\vec{x}), \hat\Psi'(\eta,\vec{y})] = - \frac{i}{a^2} \delta^{(3)}(\vec{x} - \vec{y}), \quad [\hat\Psi(\eta,\vec{x}), \hat\Psi(\eta,\vec{y})] = 0, \quad [\hat\Psi'(\eta,\vec{x}), \hat\Psi'(\eta,\vec{y})] = 0.  \label{canCommPsi} \ee
%This conditions are fulfilled provided that the annihilation and creation operators, $b_{\vec{k}}$ and $b_{\vec{k}}^\dagger$, 
%which appear in the
We can now perform an expansion in three-dimensional plane waves $e^{i \vec{q}\cdot\vec{x}}$ of the field
\be  \hat\Psi(\eta,\vec{x}) =  \int \frac{d^3q}{(2\pi)^{3/2}}e^{i \vec{q}\cdot\vec{x}}  \Psi_0(\eta,\vec{q}). \ee
%\be  \hat\Psi(\eta,\vec{x}) =  \int \frac{d^3q}{(2\pi)^{3/2}}e^{i \vec{q}\cdot\vec{x}} b_{\vec{q}} \hat\Psi_{\vec{q}}(\eta)  + \mbox{h.c.}  \ee
Notice that the hermiticity condition on $\Psi$ implies 
\be \Psi_0(\eta,\vec{q})^\dagger  = \Psi_0(\eta, - \vec{q}). \label{HermCondPsi} \ee
 The mode functions $ \hat\Psi_0$ are the solutions of the field equation   in momentum space 
  \be   \Psi''_0+2{\cal H}  \Psi'_0+(q^2-4{\cal H}^2) \Psi_0 = 0. \label{hatPsi-eq} \ee
  The two independent solutions are 
  \be g_0(\eta, q) = \frac{H|\eta|}{\sqrt{2q}} \left(1-\frac{3i}{q\eta}-\frac{3}{q^2\eta^2}\right)e^{-iq\eta} \qquad \mbox{ and its complex conjugate}. \label{PsiW} \ee
 We see that these functions reproduce the two solutions in (\ref{Psibehaviour}) up to $M_2^2/H^2$ corrections; these corrections are the effect of the Einstein-Hilbert term that here we have neglected, but is of course important to recover Einstein's gravity when $M_2 \gg H$.  
  So, by using the hermiticity condition in (\ref{HermCondPsi}), we have 
  \be  \Psi_0(\eta, \vec{q}) = b_0(\vec{q}) g_0 (\eta, q) +b_0(-\vec{q})^\dagger  g_0 (\eta, q)^*,   \ee
  where $b_0(\vec{q})$ is an operator, which  will be identified with the annihilation operator of the quanta of the scalar field under study.
  
  One can show now that   the canonical commutators (\ref{canCommPsi}) imply\footnote{The fact that these commutation relations for $b_0$ and $b_0^\dagger$ are implied by the canonical commutators of the fields (for the normalization of the modes given in (\ref{PsiW})) can be proved in the same way as we did for scalar fields: one expresses $b_0$ and $b_0^\dagger$ in terms of the field with a relation analogue to (\ref{aqExpr}) and then one uses the canonical commutators for the fields. A normalization  of the modes different from the one in (\ref{PsiW}) would have lead to a constant in front of $\delta(\vec{k}-\vec{q})$ in (\ref{bComm}) different from $-1$. }  
  \be [ b_0(\vec{k}),  b_0(\vec{q})^\dagger] = -  \delta(\vec{k}-\vec{q}), \qquad [ b_0(\vec{k}),  b_0(\vec{q})] = 0.  \label{bComm}\ee
  In appendix \ref{Hamiltonian approach} we show that, by introducing a negative norm to the states with a odd number of quanta created and annihilated by $b_0^\dagger$ and $b_0$ respectively, one obtains a Hamiltonian that is bounded from below. A possible way of addressing the issues  generated by negative norms will be discussed in section \ref{Issues due to the Weyl term}.
 The expression of the Hamiltonian  in terms of $b_0^\dagger$ and $b_0$ found in  appendix \ref{Hamiltonian approach} and the commutation relations in (\ref{bComm}) allows us to interpret them as creation and annihilation operators, respectively.  %  The minus sign here leads to the negative norm states.
%  
%  \xxx{Is the corresponding Hamiltonian bounded from below? }
  %Is the Hamiltonian of a field on de Sitter conserved? No, consider for example a simple scalar field 

\subsection{Inclusion of slow-roll} \label{Inclusion of slow-roll}

We now turn to the effect of a non-zero slow-roll. 

\subsubsection{Newtonian gauge}

We start from the curvature perturbation ${\cal R}$, which, in the Newtonian gauge, reads
\be{\cal R} = -\Psi-{\cal H} \frac{K_{ij}\phi'^i\varphi^j}{K_{lm}\phi'^l\phi'^m}. \label{Rmulti}\ee
% (see eqs. (5.4.1) at page 246 and (10.4.14) at page 499 of Weinberg's cosmology book)
 In our case $\mathcal{R}$ can be simplified by using the expression of $K_{ij}$ given in eq. (\ref{KijRen}), which leads to 
 \be{\cal R} = -\Psi -{\cal H} \frac{\sum_i\phi'^i\varphi^i}{\sum_j\phi'^j\phi'^j}. \label{Rmulti3}\ee	
As well-known, from ${\cal R}$ we can extract important observable quantities, such as its power spectrum   $P_{\cal R}(q)$ and the spectral index $n_s$. In particular we are interesting in $\mathcal{R}$ at superhorizon scales. Here we show that $\mathcal{R}$ in this case is insensitive to the metric perturbation $\Psi$, at least in the slow-roll approximation and is therefore given by its known value in the absence of the Weyl-squared term.

We start by considering the field equation for the scalar field perturbations $\varphi^i$ from eq. (\ref{SEH}): 
\bea 0 &=&  K_{ij} \varphi''^j+2{\cal H}  K_{ij} \varphi'^j +2\gamma^a_{jl} K_{ai} \phi'^l\varphi'^j  +2{\cal H} K_{ji,l} \phi'^j\varphi^l  \nonumber \\ &&+\left(K_{ji,l} \phi'^j\right)' \varphi^l -K_{ij}\vec{\nabla}^2\varphi^j+a^2 U_{,i,j} \varphi^j  \nonumber \\
&&-{\cal H}(6\Psi+2\Phi)K_{ji} \phi'^j-(3\Psi'+\Phi')K_{ji}\phi'^j-(3\Psi+\Phi)\left(K_{ji}\phi'^j\right)' \nonumber \\ &&+ \frac{3\Psi+\Phi}{2}K_{lj,i}\phi'^l\phi'^j+a^2(\Phi-3\Psi) U_{,i}. \nonumber
\eea
In the leading non-trivial slow-roll approximation we can approximate $\Phi=- \Psi$, which was derived in the pure de Sitter case, because $\Phi$ and $\Psi$ always appear multiplied by some derivative of background quantities and therefore we can use for them the zero-order slow-roll approximation. We can also use the  slow-roll approximations $\phi'^i\phi'^j \approx 0$, $\phi''^i\approx {\cal H} \phi'^i$ and the scalar field equation (\ref{ScFiEq}). By multiplying by $K^{im}$ and summing over $i$ we then obtain
\be \varphi''^m +2{\cal H} \varphi'^m+2 \gamma^m_{jl} \phi'^l\varphi'^j +3{\cal H} K^{im} K_{ji,l}\phi'^j \varphi^l -\vec{\nabla}^2 \varphi^m +a^2 K^{im} U_{,i,j} \varphi^j = 2\phi'^m \Psi' + 2a^2 U^{,m} \Psi.\ee

We now restrict the analysis to the scalar field perturbations that are effectively massless. Indeed, as shown in the zero-order slow-roll approximation in section \ref{Pure de Sitter}, only those fields contribute in the superhorizon case. This means that we can neglect the term $a^2 K^{im} U_{,i,j} \varphi^j $ in the equation above, which can now be written 
\be \varphi''^m +2{\cal H} \varphi'^m -\vec{\nabla}^2 \varphi^m  = 2\phi'^m \Psi' + 2a^2 U^{,m} \Psi-2 \gamma^m_{jl} \phi'^l\varphi'^j -3{\cal H} K^{im} K_{ji,l}\phi'^j \varphi^l. \label{EqVarphii}\ee
Since we want to find the solution $\varphi^m$ in the next-to-leading order in the slow-roll approximation we can substitute in the right-hand-side of this equation the values for $\Psi$ and $\varphi^i$ that we have found in section  \ref{Pure de Sitter} in the pure de Sitter case; we refer to these quantities as $\Psi_{\rm dS}$ and $\varphi^i_{\rm dS}$. Eq. (\ref{EqVarphii}) then becomes an inhomogeneous equation with the following source term 
\be 2\phi'^m \Psi_{\rm dS}' + 2a^2 U^{,m} \Psi_{\rm dS}-2 \gamma^m_{jl} \phi'^l\varphi'^j_{\rm dS} 
-3{\cal H} K^{im} K_{ji,l}\phi'^j \varphi_{\rm dS}^l \approx 2\phi'^m \Psi_{\rm dS}' + 2a^2 U^{,m} \Psi_{\rm dS}, 
\ee
where in the latter approximation we have used that $\gamma^m_{jl} \phi'^l\varphi'^j_{\rm dS}$ goes as $a$ and $3{\cal H} K^{im} K_{ji,l}\phi'^j \varphi_{\rm dS}^l$ as $a^2$ in the superhorizon limit (having used $\phi'^i \approx -a^2 U^{,i}/(3{\cal H})$ and $\varphi'^j, \varphi^j \sim constant$ in that limit), while the other terms $2\phi'^m \Psi_{\rm dS}' + 2a^2 U^{,m} \Psi_{\rm dS}$ go as $a^3$ (having used the behaviour of $\Psi$ given in eq. (\ref{Psibehaviour}) for a moderate ghost mass\footnote{The opposite limit $M_2 \gg H$ corresponds to the limit in which Einstein theory is recovered, together with its prediction for ${\cal R}$.}, $M_2 \lesssim H$). Eq. (\ref{EqVarphii}) then becomes 
\be \varphi''^m +2{\cal H} \varphi'^m -\vec{\nabla}^2 \varphi^m  = 2\phi'^m \Psi'_{\rm dS} + 2a^2 U^{,m} \Psi_{\rm dS}. \label{EqVarphii2}\ee
Moreover, notice that ${\cal H}$ appearing in the second term of the left-hand-side of this equation can be replaced by the corresponding quantity in the pure de Sitter case: indeed, eq. (\ref{EOM2c}) shows that the difference between the pure de Sitter $\mathcal{H}$ and the space-time that takes into account the dynamics of the scalar fields is beyond the next-to-leading order slow-roll approximation that we are using here.
%taking the time-derivative of eq. (\ref{EOM1c}) one finds that ${\cal H}'$ .

By using now the slow-roll equations $\phi''^i \approx {\cal H} \phi'^i$, $\phi'''^i \approx 2{\cal H}^2 \phi'^i$ and ${\cal H}' \approx {\cal H}^2$, as well as the  equations of motion (\ref{EOMvarphi}) and (\ref{hatPsi-eq}), one finds that the following configuration is a solution in the next-to-leading slow-roll approximation
\be \varphi^i =\varphi^i_{\rm dS} -\frac{\phi'^i}{{\cal H}} \Psi_{\rm dS}. 
\ee

By using this solution in eq. (\ref{Rmulti3}) we find 
\be{\cal R} = -\Psi_{\rm dS} -{\cal H} \frac{\sum_i\phi'^i\varphi^i}{\sum_j\phi'^j\phi'^j} =  -{\cal H} \frac{\sum_i\phi'^i\varphi_{\rm dS}^i}{\sum_j\phi'^j\phi'^j}, \label{Rmulti2}\ee
where we have substituted $\Psi \rightarrow \Psi_{\rm dS}$ in the first term of ${\cal R}$ because in the second term we have two time-derivatives in the denominator and only one in the numerator and thus going to the next-to-leading order in the slow-roll approximation in $\varphi^i$ corresponds to the zero-order approximation in the first term of ${\cal R}$. We see that the dependence on the ghost cancels in ${\cal R}$, which therefore reproduces the expression that we have when gravity is described by Einstein's theory coupled to an arbitrary number of scalar fields. The corresponding power spectrum and the spectral index $n_s$  will be recalled in section \ref{Observational quantities}.

\subsubsection{Co-moving gauge}

We conclude this section by showing that the superhorizon divergence of $\Psi$ is a gauge artifact. We extend the validity of a previous argument of \cite{Ivanov:2016hcm} to theories with a generic number of scalar fields. 

To do so we consider the co-moving gauge defined by\footnote{ Even in the multifield case, the energy momentum tensor up to the linear level in the perturbations has the form of a perfect fluid $T_{\mu\nu} = p g_{\mu\nu} + (p+\rho) u^\mu u^\nu$, where $p$ is the pressure, $\rho$ is the energy density and $u^\mu$ is the four-velocity of the perfect fluid (see e.g. \cite{Weinberg:2008zzc}). We define $\delta u$ as the scalar component of the spatial part of the four-velocity, that is  $\delta u_i = \partial_i \delta u + ...$, where the dots represent a divergenceless spatial vector.
}
  $\delta u = 0$.  Given that generically
  \be \delta u = - a \frac{K_{ij} \phi'^i \varphi^j}{K_{lm} \phi'^l \phi'^m},\ee
  setting $\delta u = 0$ with a gauge transformation starting from the Newtonian gauge produces a non diagonal metric for the scalar perturbations of the form 
 \be ds^2_{\rm scalar}= a^2 (1+2A) d\eta^2 -2 a \partial_i B  d\eta dx^i  -a^2 (1+2\mathcal{R}) \delta_{ij} dx^idx^j, \label{Co-movMet}\ee
 where $\mathcal{R}$ is the curvature perturbation defined in (\ref{Rmulti}),
 \be B= \frac{K_{ij} \phi'^i \varphi^j}{K_{lm} \phi'^l \phi'^m} \ee
and $A$ is obtained from the Newtonian potential $\Phi$ by adding a term proportional to the time-derivative of $aB$ and thus does not represent another independent degree of freedom.

We have just shown that $\mathcal{R}$ is not sensitive to the superhorizon divergence of $\Psi$. As far as $B$ is concerned, we observe that eq. (\ref{Rmulti3}) tells us 
\be B = -\frac1{\mathcal{H}} \mathcal{R} -\frac1{\mathcal{H}} \Psi.\ee
The first   term in $B$ go to zero at superhorizon scales. Regarding the second one we have to distinguish between two situations. 
\begin{itemize}
\item $M_2^2/H^2 \gtrsim 1/N$. In this case also $\Psi/\mathcal{H}$ is suppressed at superhorizon scales, $\eta \sim e^{-N}\ll  1$, where we have taken into account the contribution of the Einstein-Hilbert term to the superhorizon behaviour of $\Psi$, eq. (\ref{Psibehaviour}).  
\item $M_2^2/H^2 \lesssim 1/N$. In this case the behaviours in (\ref{Psibehaviour})  practically reduce to $1/\eta$ and $1$, meaning that $B$ does not vanish at superhorizon scales. Also in this situation, however, $B$ remains finite because $\mathcal{H} \approx -1/\eta$.%, as a consequence of  (\ref{PsiW}).
\end{itemize}
Ref.  \cite{Ivanov:2016hcm} did not consider the first case. Here we include it to keep the analysis general.    Therefore, we conclude not only that the divergence of $\Psi$ is a gauge artifact, but also that  this extra degree of freedom due to the ghost is either suppressed in the superhorizon limit  (for $M_2^2/H^2 \gtrsim 1/N$) or remains finite (for $M_2^2/H^2 \lesssim 1/N$).

\section{Vector perturbations}\label{Vector perturbations}

 The contribution of $S_{ES}$ to the quadratic action for the tensor perturbations is 
 \be S_{ES}^{(V)} = \frac{\bp^2}{4} \int d^4x \, a^2\left(\partial_i V_{j}\right)^2,\ee
 where the space indices are contracted with the flat metric  $\delta_{ij}$. The contribution of $S_W$  is instead
 \be S^{(V)}_W =- \frac{1}{2f_2^2} \int d^4x \left(\partial_i V_j' \partial_i V_j' - V_i \vec{\nabla}^4 V_i\right). \ee
We explicitly checked that the presence of an arbitrary number of scalar fields does not change the form of $S_{ES}^{(V)}$ and $S^{(V)}_W$. In the case of $S_{ES}^{(V)}$ the check requires the use of the background equations (\ref{EOM1c})-(\ref{EOM2c}).
 
 Thus the full action for vector perturbations $S_V = S_{ES}^{(V)} + S^{(V)}_W$ is given by  $S_V =\int d^4x  \Lag_{V}$ where 
\be \Lag_{V} =  \frac{\bp^2}{4}\left[-a^2 V_{j}\vec{\nabla}^2 V_{j} + \frac1{M_2^2}\left( V_j' \vec{\nabla}^2 V_j' + V_i \vec{\nabla}^4 V_i\right)\right].  \label{LagV}\ee 
We observe that this action does not contain terms with four time-derivatives. This follows from the fact that the Einstein-Hilbert action (plus an arbitrary number of scalar fields) does not lead to any propagating  field with helicity 1 or -1, while the massive ghost should provide two fields, one with helicity 1 and the other one with helicity -1. The $V_i$ account for these two fields. If we had found terms with four time-derivatives for the $V_i$ we would have a number of degrees of freedom doubled with respect to the correct one.

We now introduce the conjugate momenta
\be P_i = \frac{\partial \Lag}{\partial V_i'} = \frac{\bp^2}{2M_2^2} \vec{\nabla}^2 V_i'. \label{PiDef} \ee
%quantize this system with the canonical commutators. 
The Laplace operator $\vec{\nabla}^2$ in the expression of the conjugate momenta was missed in the previous studies of vector perturbations of \cite{Myung:2014jha}. As we will see, the presence of such operator modify the momentum dependence of the vector modes and the commutation rules of the creation and annihilation operators.

The quantization is obtained as usual by imposing the canonical commutators. In order to do so, however, one should identify the independent degrees of freedom. The condition in (\ref{Cond1}) tells us that for a plane wave with given momentum $\vec{q}$ there are only two independent components. So to identify the two independent degrees of freedom we go to momentum space and write for
 $F_j \equiv\{V_j, P_j \}$ 
\bea F_{j}(\eta, \vec{x}) =  \int \frac{d^3q}{(2\pi)^{3/2}}  e^{i\vec{q}\cdot \vec{x}} \sum_{\lambda= \pm 1}  F_\lambda(\eta,\vec{q}) e^\lambda_{j} (\hat q),  \label{ExpMomV}  \eea
where $e^\lambda_{j} (\hat q)$ are the usual polarization vectors for helicities $\lambda= \pm 1$. We recall that for $\hat q$ along the third axis the polarization tensors that satisfy (\ref{Cond1}) are  given by 
\be e^{+1}_{1}  = 1/\sqrt2, \quad e^{+1}_{2} = i/\sqrt2, \quad e^{+1}_{3} = 0, \quad  e^{-1}_{j} =  (e^{+1}_{j})^* \label{PolV}\ee 
and for a generic momentum direction $\hat q$ we can obtain $e^\lambda_{j} (\hat q)$ by applying to (\ref{PolV}) a rotation that connects the third axis with  $\hat q$.
The polarization vectors defined in this way obey 
\be e^\lambda_{j} (\hat q) (e^{\lambda'}_{j} (\hat q))^* = \delta^{\lambda\lambda'}.  \label{PolVecOrt}\ee 
We can now impose the canonical commutators:
\be [V_\lambda(\eta, \vec{q}), (P_{\lambda'}(\eta, \vec{k}))^\dagger] = i \delta_{\lambda\lambda'}  \delta^{(3)}(\vec{q}-\vec{k}), \qquad \mbox{(and all the other commutators vanishing),}
% \mbox{(and all the other commutators vanishing)} 
\label{CCmomV} \ee
which, according to the expansion in (\ref{ExpMomV}) and the condition in (\ref{PolVecOrt}), lead to the following canonical commutators in coordinate space:
\be  [V_{j}(\eta, \vec{x}), P_j(\eta, \vec{y})] = 2i  \delta^{(3)}(\vec{x}-\vec{y}), \qquad \mbox{(and all the other commutators vanishing)}. \ee 

%By using the Hamilton equations
%\be V_\lambda' = \frac{\partial H_V}{\partial P_\lambda^*}, \qquad P_\lambda' = -  \frac{\partial H_V}{\partial p_\lambda^*} \ee
By taking the variation of the action we obtain the equations of motion for the vector perturbations, which, working in momentum space, read
\be V_\lambda'' = -\left(q^2+M_2^2 a^2\right) V_\lambda.  \label{EqVpert}\ee

We now solve these equation in the pure de Sitter case, for which $a^2(\eta) = 1/(H^2\eta^2)$. Indeed (\ref{EOM2c}) shows that the error that is produced in this way is beyond the next-to-leading order slow-roll approximation that we are using here.  By defining $z \equiv - q\eta$ and $\rho \equiv H^2/M_2^2$ we find 
\be \frac{d^2V_\lambda}{dz^2} +\left(1+\frac1{\rho z^2}\right)V_\lambda =0,\ee
whose linearly independent solutions are\footnote{The label 1 on $f_1$ has been introduced to emphasise that we are in the helicity $\pm 1$ sector.}
\be  f_1(z)\equiv   \sqrt{z} J_{\frac{\sqrt{\rho -4}}{2 \sqrt{\rho }}}(z)+i \sqrt{z} Y_{\frac{\sqrt{\rho -4}}{2 \sqrt{\rho }}}(z) \qquad \mbox{ and its complex conjugate.} \ee

We can therefore expand
\be V_\lambda(\eta,\vec{q}) = \gamma_\lambda(\vec{q}) f_1(-q \eta) +\gamma^\dagger_{-\lambda}(-\vec{q}) f^*_1(-q \eta), \ee
where we have used the reality condition $V_j(\eta,\vec{x})^\dagger= V_j(\eta,\vec{x})$ for the fields in coordinate space, which corresponds to $V_\lambda(\eta,\vec{q})^\dagger= V_{-\lambda}(\eta, -\vec{q})$ in momentum space. The quantities   $ \gamma_\lambda(\vec{q})$ are to be interpreted as operators in the quantum theory, but their normalization  is not fixed. From the analysis of the scalar sector we have learned  that  the way to properly normalize them is to impose the canonical commutators and require that the $\gamma_\lambda$ together with their hermitian conjugate satisfy the commutation rules for creation and annihilation operators. To achieve this goal we observe that the $\gamma_\lambda$ can be expressed as a functional of $V_i$ with a relation analogue to (\ref{aqExpr}). This shows that there is only one assignment for the commutation rules satisfied by  $\gamma_\lambda$. This assignment turns out to be
\be [\gamma_\lambda(\vec{q}), \gamma_\lambda^\dagger(\vec{k})] = -   \frac{c_\gamma(q)}{\mathcal{F}\mathcal{F}^*}  \delta_{\lambda \lambda'}\delta^{(3)}(\vec{q}-\vec{k}), \label{CCccd}\ee
and all the other commutators equal to zero, where 
\be  c_\gamma(q) \equiv \frac{M_2^2}{\bp^2 q^3} \qquad \mbox{and} \qquad  \mathcal{F} \equiv  \frac{(1-i) e^{-\frac{1}{4} i \pi  \sqrt{\frac{\rho -4}{\rho }}}}{\sqrt{\pi }}.\ee 
Notice that without knowing $\rho$ we cannot simplify $\mathcal{F}\mathcal{F}^*$ as we leave open the possibility that $\rho < 4$.
These commutators can be brought into the more standard form 
\be [c_\lambda(\vec{q}), c_\lambda^\dagger(\vec{k})] = -  \delta_{\lambda \lambda'}\delta^{(3)}(\vec{q}-\vec{k}) \label{CCccd2}\ee
by defining 
\be c_\lambda(\vec{q}) \equiv  \frac{\mathcal{F}}{\sqrt{c_\gamma(q)}} \gamma_\lambda(\vec{q}), 
 \ee
 which also leads to the properly normalized modes 
 \be    g_1(\eta,q) =   \frac{\sqrt{\pi} M_2e^{i\frac{\pi}4\left(1+\sqrt{1-4\frac{M^2_2}{H^2}}\right) }}{\sqrt{2}\bp q^{3/2}} \sqrt{-q\eta}\left(J_{\sqrt{\frac14 -\frac{M_2^2}{H^2}}}(-q\eta) + i Y_{\sqrt{\frac14 -\frac{M_2^2}{H^2}}}(-q\eta) \right) \label{ModFc} \ee
 such that the initial function can be expressed  as follows
\be V_\lambda(\eta, \vec{q}) = c_\lambda(\vec{q}) g_1(\eta, q) + c^\dagger _{-\lambda}(-\vec{q}) g^*_1(\eta, q).\ee
In a  previous calculation  ref. \cite{Myung:2014jha} found the opposite result for the commutator in  (\ref{CCccd2}).  This difference is due to the fact that we took into account the operator $\vec{\nabla}^2$ in the definition of the conjugate momenta, eq. (\ref{PiDef}), which effectively changes the overall sign when going to momentum space: $\vec{\nabla}^2\rightarrow -q^2$. 

% corresponds to negative norm quanta.  This fact was missed in the previous calculation in \cite{Myung:2014jha} were positive norm commutation rules were found. However, the presence of  negative norm quanta in this sector is suggested by the fact that, as we have stated before, $V_i$ account for the  helicity 1 and the   helicity -1 component  of the massive ghost.

The expression of $g_1$ we find differs  from  the previous determinations of ref. \cite{Myung:2014jha}: we have a factor of $q^{3/2}$ in the denominator, instead of $q^{1/2}$; this difference is also due to the fact that we took into account the operator $\vec{\nabla}^2$ in the definition of the conjugate momenta.
% Second, we have the factor
%\be e^{i\frac{\pi}8\left(\sqrt{1-4\frac{M^2_2}{H^2}}-\sqrt{1-4\frac{M^2_2}{H^2}}^*\right) } 
%= \left\{ \ba {c} e^{-\frac{\pi}4\sqrt{4\frac{M^2_2}{H^2}-1}},
% \quad \mbox{for} \quad M_2>2 H\\  \\1, \quad \mbox{for} \quad M_2<2 H \ea \right.
%\label{expCR}\ee 
%which agrees with the result in \cite{Myung:2014jha}  only for  $M_2>2H$. 
We also observe that for  $M_2>H/2$ a complex exponential  appearing in $g_1$ becomes real,
\be e^{i\frac{\pi}4\sqrt{1-4\frac{M^2_2}{H^2}} }  = e^{-\frac{\pi}4\sqrt{4\frac{M^2_2}{H^2}-1} }, \qquad  (M_2>2H), \ee 
and exponentially suppresses the ghost mode $g_1$ for $M_2 \gg H$. This is what we expect because for $M_2 \gg H$ the effect of the ghost on the inflationary perturbations should disappear. Notice, moreover, that for any value of $M_2$ the vector modes $g_1$ are suppressed\footnote{ To see this one can use the expansions of the Bessel functions for small $z$:
\bea J_ n(z) &=&  z^n \left(\frac{2^{-n}}{\Gamma (n+1)}+O\left(z^2\right)\right), \\  Y_n(z) &=& z^{-n} \left(-\frac{2^n \Gamma (n)}{\pi }+O\left(z^2\right)\right)+z^n \left(-\frac{2^{-n} \cos (n \pi )\Gamma (-n)}{\pi }+O\left(z^2\right)\right). \eea
Notice that the superhorizon limit consists in taking $\eta \sim e^{-N}$, but even if $M_2^2/H^2 \lesssim 1/N$, we would have a suppression of order $M_2/H$ (compared to the scalar modes in (\ref{ModeVarphi})) due to the overall coefficient in  (\ref{ModFc}).} at superhorizon scales.

We have seen from the analysis of the scalar perturbations that the commutation rules in (\ref{CCccd2}) are not enough to identify $c_\lambda$ and $c_\lambda^\dagger$ as annihilation and creation operators, respectively, but one should see how these operators appear in the Hamiltonian. This is done in 
 appendix \ref{Vector perturbations app} where this identification is justified and it is also shown that the Hamiltonian does not have negative eigenvalues, at least when it is conserved, if one introduces a negative norm for the states with an odd number of quanta created by $c_\lambda^\dagger$.

\section{Tensor perturbations} \label{Tensor perturbations}

The contribution of $S_{ES}$ to the quadratic action for the tensor perturbations is well-known (see for example the text-book \cite{Mukhanov:2005sc})
\be S_{ES}^{(T)} = \frac{\bp^2}{8} \int d^4x \, a^2\left(h'_{ij}h'_{ij}+h_{ij}\vec{\nabla}^2h_{ij}\right).\ee
The contribution of $S_W$ is instead 
\be S^{(T)}_W =- \frac{1}{4f_2^2} \int d^4x \left( h''_{ij} h''_{ij}+2 h'_{ij}\vec{\nabla}^2 h'_{ij} + h_{ij}\vec{\nabla}^4 h_{ij}\right).\ee
One can explicitly check that the presence of an arbitrary number of scalar fields do not change the form of $S_{ES}^{(T)}$ and $S^{(T)}_W$; the check for $S_{ES}^{(T)}$  requires the use of the background equations (\ref{EOM1c})-(\ref{EOM2c}).

The quadratic action for the tensor perturbations $S_{T} =S^{(T)}_W  +S^{(T)}_{EH}$ can be written in terms of the Lagrangian in the usual way  $ S_{T}  = \int d^4x  \Lag_{T},$ where 
\be   \Lag_{T} =   \frac{\bp^2a^2}{8} \left(h'_{ij}h'_{ij}+h_{ij}\vec{\nabla}^2h_{ij}\right)- \frac{\bp^2}{8M_2^2} \left( h''_{ij} h''_{ij}+2 h'_{ij}\vec{\nabla}^2 h'_{ij} + h_{ij}\vec{\nabla}^4 h_{ij}\right). \label{Stensor}\ee

We now introduce the canonical formalism that is suitable to quantize the system. To do so we use the Ostrogradsky canonical method \cite{Ostrogradsky,Pais}; this method was introduced for Lagrangians without explicit dependence on time, but can be applied without significant modifications in the present case where the Lagrangian does have such dependence (due to the cosmological scale factor), as explained in appendix \ref{Lagrange and Hamilton}. 
This system with four derivatives can be transformed into one with two derivatives by doubling the degrees of freedom 
\be h_{ij}^{(1)} = h_{ij}, \qquad  h_{ij}^{(2)} = h'_{ij}.  \label{canCoord}\ee
We can define the conjugate variables as follows
\be p^{(1)}_{ij} = \frac{\delta \Lag}{\delta h_{ij}'^{(1)}}, \qquad p^{(2)}_{ij} = \frac{\delta \Lag}{\delta h_{ij}'^{(2)}},  
\ee
where we have introduced the variational derivatives for a generic variable $X$
\be \frac{\delta\Lag}{\delta X} = \frac{\partial\Lag}{\partial X} - \frac{d}{dt} \frac{\partial \Lag}{\partial \dot X}, 
\ee 
which generalizes the usual formula to four-derivative theories. We obtain
\be p^{(1)}_{ij} = \frac{\bp^2}{4}\left(a^2 h'_{ij} -\frac{2}{M_2^2}\vec{\nabla}^2h'_{ij}  +\frac{1}{M_2^2} h'''_{ij} \right) , \qquad p^{(2)}_{ij} =-\frac{\bp^2}{4M_2^2}  h''_{ij}. \label{CanMom} \ee

 The quantization is obtained as usual by imposing the canonical commutators. In order to do so, however, one should identify the independent degrees of freedom. The conditions in (\ref{Cond2}) tell us that for a plane wave with given momentum $\vec{q}$ there are only two independent components for each canonical variable. So to identify the two  independent degrees of freedom we go to momentum space and write for
 $F_{ij} \equiv \{ h^{(1)}_{ij}, p^{(1)}_{ij}, h^{(2)}_{ij}, p^{(1)}_{ij} \}$, %with $l=1,2$,
\bea F_{ij}(\eta, \vec{x}) =  \int \frac{d^3q}{(2\pi)^{3/2}}  e^{i\vec{q}\cdot \vec{x}} \sum_{\lambda= \pm 2}  F_\lambda(\eta,\vec{q}) e^\lambda_{ij} (\hat q),  \label{ExpMom}  \eea
where $e^\lambda_{ij} (\hat q)$ are the usual polarization tensors for helicities $\lambda= \pm 2$. We recall that for $\hat q$ along the third axis the polarization tensors that satisfy (\ref{Cond2}) are  given by 
\be e^{+2}_{11} = -e^{+2}_{22} = 1/2, \quad e^{+2}_{12} = e^{+2}_{21} = i/2, \quad e^{+2}_{3i} =e^{+2}_{i3} = 0, \quad  e^{-2}_{ij} =  (e^{+2}_{ij})^* \label{PolT}\ee 
and for a generic momentum direction $\hat q$ we can obtain $e^\lambda_{ij} (\hat q)$ by applying to (\ref{PolT}) a rotation that connects the third axis with  $\hat q$.
The polarization tensors defined in this way obey 
\be e^\lambda_{ij} (\hat q) (e^{\lambda'}_{ij} (\hat q))^* = \delta^{\lambda\lambda'}.  \label{PolOrt}\ee 
As expected the tensor sector includes two fields with helicities $\pm 2$: they correspond to the graviton and the helicity $\pm 2$ components of the spin-2 ghost. These two fields make together the field $h_{ij}$ with a four-derivative Lagrangian.
We can now impose the canonical commutators to the variable $F_{\lambda}$:
\be\hspace{-0.2cm} [h^{(l)}_\lambda(\eta, \vec{q}), p^{(l')}_{\lambda'}(\eta, \vec{k})^\dagger] = i \delta_{\lambda\lambda'} \delta^{l l'}\delta^{(3)}(\vec{q}-\vec{k}),  \qquad \mbox{(and all the other commutators vanishing),}
%\qquad \mbox{(and all the other commutators vanishing)} 
\label{CCmom} \ee
which, according to the expansion in (\ref{ExpMom}) and the condition in (\ref{PolOrt}), lead to the following canonical commutators in coordinate space:
\be  [h^{(l)}_{ij}(\eta, \vec{x}), p^{(l')}_{ij}(\eta, \vec{y})] = 2i \delta^{l l'}\delta^{(3)}(\vec{x}-\vec{y}). \qquad \mbox{(and all the other commutators vanishing)}. 
\ee 

Now that we know how to quantize we come back to the action in (\ref{Stensor}) and the associated equation 
\be (a^2h'_{ij})' - a^2 \vec{\nabla}^2 h_{ij} +\frac1{M_2^2} \left(h''''_{ij} -2\vec{\nabla}^2 h''_{ij}+\vec{\nabla}^4 h_{ij}\right) = 0. \ee
By using the expansion in (\ref{ExpMom}) for $h_{ij}$ we obtain 
\be  (a^2h'_\lambda)' + a^2q^2 h_\lambda +\frac1{M_2^2} \left(h''''_\lambda +2q^2 h''_\lambda+q^4 h_\lambda\right) = 0. \ee
notice that, again, we can replace  $a^2$ and $\mathcal{H}$  here with the their pure de Sitter expressions, $a^2(\eta)= 1/(H^2\eta^2)$ and $\mathcal{H} = -1/\eta$: indeed (\ref{EOM2c}) shows that the error that is produced in this way is beyond the next-to-leading order slow-roll approximation that we are using here. We then obtain
\be \frac{d^4}{dz^4} h_\lambda+2\frac{d^2}{dz^2} h_\lambda + h_\lambda +  \frac1{\rho} \left[\frac{d}{dz}\left(\frac1{z^2} \frac{d}{dz} h_\lambda\right) +\frac1{z^2} h_\lambda\right]  =0, \label{4derEq}  \ee
where we have introduced again $z\equiv -q \eta$ and $\rho \equiv H^2/M_2^2$.
We now follow the method of  \cite{Clunan:2009er,Deruelle:2012xv} to solve this equation and improve it, showing that it provides all the linearly independent solutions. We write the differential operator appearing in this equation, 
\be \mathcal{D}_z  \equiv \frac{d^4}{dz^4} +2\frac{d^2}{dz^2} + 1 + \frac1{\rho} \left(\frac1{z^2}\frac{d^2}{dz^2} -\frac2{z^3}\frac{d}{dz} +\frac1{z^2}\right),   \ee
in two equivalent ways 
\bea \mathcal{D}_z &=& \left( \frac{d^2}{dz^2} +\frac2{z}\frac{d}{dz}+1+\frac1{\rho z^2}\right)  \left(\frac{d^2}{dz^2}-\frac2{z}\frac{d}{dz} +1\right),  \\ 
 \mathcal{D}_z &=&  \left( \frac1{z^2}\frac{d^2}{dz^2} -\frac2{z^3}\frac{d}{dz}+\frac1{ z^2}\right)  \left(z^2\frac{d^2}{dz^2}-2z\frac{d}{dz} +2+z^2+\frac1{\rho}\right).\eea
Therefore, the solutions of the two second order equations 
\be \left(\frac{d^2}{dz^2}-\frac2{z}\frac{d}{dz} +1\right)h_\lambda =0, \qquad \left(z^2\frac{d^2}{dz^2}-2z\frac{d}{dz} +2+z^2+\frac1{\rho}\right)h_\lambda =0\ee
are all solutions of the four-derivative equation in (\ref{4derEq}); we also show now that they are all linearly independent. Substituting $h_\lambda \rightarrow z \mu^{(1)}_\lambda$ in the first equation and $h_\lambda \rightarrow z \mu^{(2)}_\lambda$ in the second one we obtain the equivalent equations 
\be \frac{d^2\mu^{(1)}_\lambda}{dz^2} +\left(1-\frac2{z^2}\right)\mu^{(1)}_\lambda =0, \qquad  \frac{d^2\mu^{(2)}_\lambda}{dz^2} +\left(1+\frac1{\rho z^2}\right)\mu^{(2)}_\lambda =0.\ee
The two linearly independent solutions of the first equation are 
\be % \mu_\lambda^{(1)}(\eta,\vec{q}) = 
 \frac{1- i z}{z} e^{iz} \qquad \mbox{ and its complex conjugate.} \ee
While the two linearly independent solutions of the second one are 
\be \sqrt{z} J_{\frac{\sqrt{\rho -4}}{2 \sqrt{\rho }}}(z)+i \sqrt{z} Y_{\frac{\sqrt{\rho -4}}{2 \sqrt{\rho }}}(z) \qquad \mbox{ and its complex conjugate.} \ee
%where $J_n(z)$ and $Y_n(z)$ are the Bessel functions of the first and second kind, respectively.

\begin{figure}[t]
\begin{center}
  \includegraphics[scale=0.45]{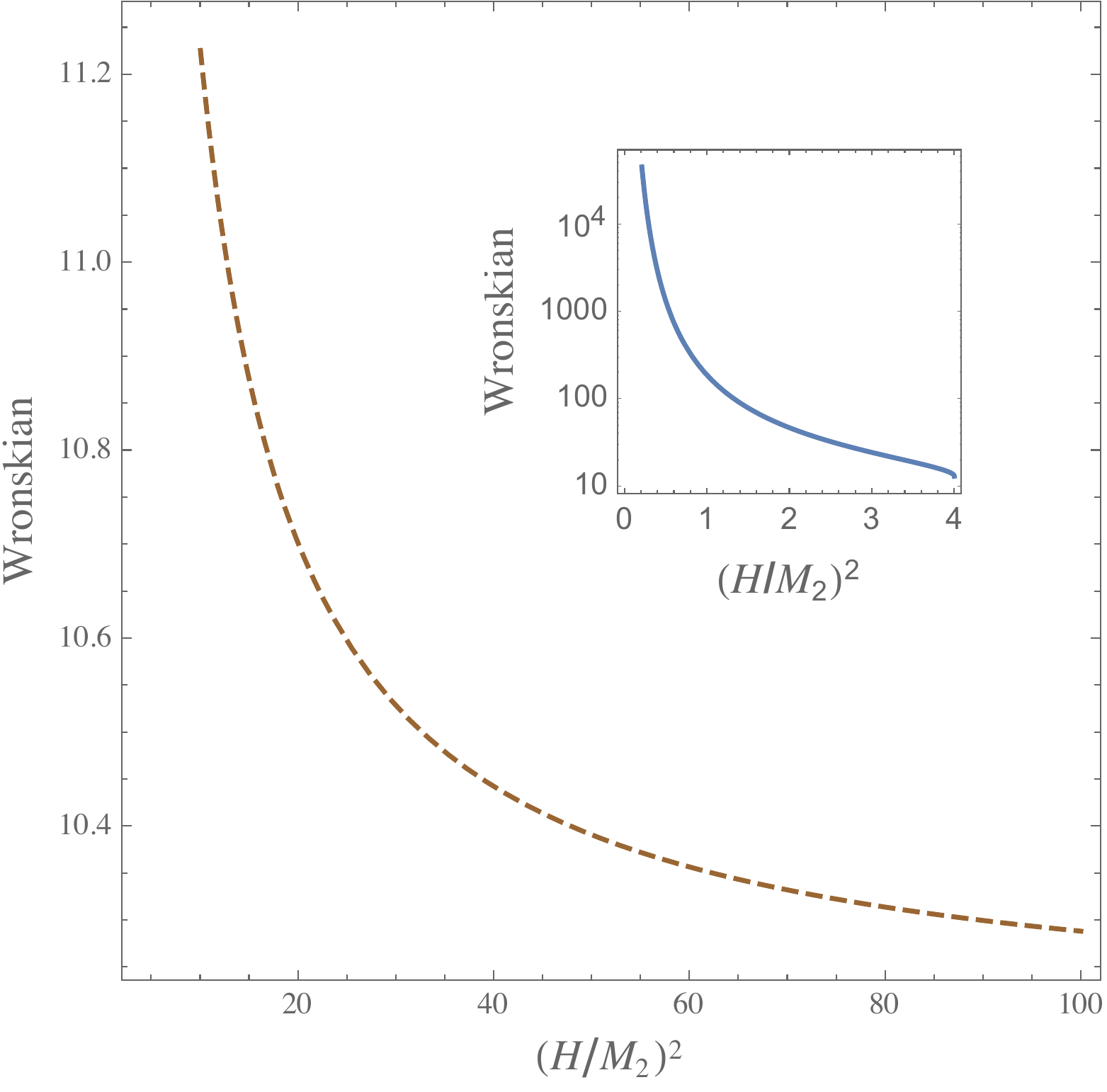} 
  \end{center}
   \caption{\em  The Wronskian of the four solutions of the tensor perturbations on de Sitter space, defined in eq. (\ref{wronskian}) as a function of the ratio between the Hubble rate and the ghost mass, showing that the solutions are linearly independent. The inset shows the behaviour for $H< 2 M_2$ with a logarithmic vertical scale. The Wronskian does not depend on $z$ because the fourth order equation (\ref{4derEq}) does not have the term with three derivatives (Liouville theorem).}
\label{wronskianF}
\end{figure}

Now, given the hermiticity condition of the field in momentum space, $h_\lambda(\eta,\vec{q})^\dagger=h_{-\lambda}(\eta, -\vec{q})$ we can write\footnote{The label $2$ on these functions reminds us that we are dealing with the helicity $\pm 2$ sector.}
\be h_\lambda(\eta, \vec{q}) = \alpha_\lambda(\vec{q}) w_2(-\eta q) +\beta_\lambda(\vec{q}) f_2(-\eta q) + \alpha^\dagger _{-\lambda}(-\vec{q}) w^*_2(-\eta q)+ \beta^\dagger _{-\lambda}(-\vec{q}) f^*_2(-\eta q),\ee
where 
\be w_2(z)\equiv (1-i z) e^{i z}, \qquad f_2(z) \equiv z^{3/2}\left(J_{\frac{\sqrt{\rho -4}}{2 \sqrt{\rho }}}(z) +iY_{\frac{\sqrt{\rho -4}}{2 \sqrt{\rho }}}(z) \right) \ee
and $\alpha_\lambda(\vec{q})$  and $\beta_\lambda(\vec{q})$ are suitable coefficient (to be interpreted as operators in the quantum theory). Their normalization can be fixed with a method similar to the one used for scalar and vector perturbations. However, the situation here is more complicated because we have four functions instead of two.  We find that these four functions  $w_2$, $w_2^*$, $f_2$, $f_2^*$ are linearly independent: their Wronskian, 
\be \mbox{Wronskian} =  \left|
\ba {cccc} w_2 & w_2^* & f_2 & f_2^*  \\
w_2' & w_2'^* & f_2' & f_2'^*  \\
w_2'' & w_2''^* & f_2'' & f_2''^*  \\
w_2''' & w_2'''^* & f_2''' & f_2'''^*    \ea \right|,  \label{wronskian} \ee
is never zero as shown in fig. \ref{wronskianF}. It follows that we can always express $\alpha_\lambda(\vec{q}), \beta_\lambda(\vec{q}), \alpha^\dagger _{-\lambda}(-\vec{q}),  \beta^\dagger _{-\lambda}(-\vec{q})$ as linear functionals of $h_\lambda(\eta, \vec{q})$. Therefore, there exists only one assignment of the commutation rules of these four operators that satisfy the canonical commutators in (\ref{CCmom}) (this result eliminates a loophole of previous determinations where the uniqueness of the commutation rules of $\alpha_\lambda(\vec{q}), \beta_\lambda(\vec{q}), \alpha^\dagger _{-\lambda}(-\vec{q}),  \beta^\dagger _{-\lambda}(-\vec{q})$ was not proved). 
%(f_\alpha, f_\alpha^*, f_\beta, f_\beta^*)
The    commutation rules we find are 
\be [\alpha_\lambda(\vec{q}), \alpha_{\lambda'}^\dagger(\vec{k})] = c_\alpha(q)  \delta_{\lambda \lambda'}\delta^{(3)}(\vec{q}-\vec{k}),\qquad  [\beta_\lambda(\vec{q}), \beta_{\lambda'}^\dagger(\vec{k})] = - \frac{c_\alpha}{\mathcal{F}^*\mathcal{F}}  \delta_{\lambda \lambda'}\delta^{(3)}(\vec{q}-\vec{k}) \label{CCaad}\ee
and all the other commutators equal to zero, where 
\be  c_\alpha(q) \equiv \frac{2H^2}{\bp^2 q^3(1+2\rho)}, \qquad \mathcal{F} \equiv  \frac{(1-i) e^{-\frac{1}{4} i \pi  \sqrt{\frac{\rho -4}{\rho }}}}{\sqrt{\pi }}.\ee 

These commutators can be brought into the more standard form 
\be [a_\lambda(\vec{q}), a_{\lambda'}^\dagger(\vec{k})] = \delta_{\lambda \lambda'}\delta^{(3)}(\vec{q}-\vec{k}),\qquad  [b_\lambda(\vec{q}), b_{\lambda'}^\dagger(\vec{k})] = -   \delta_{\lambda \lambda'}\delta^{(3)}(\vec{q}-\vec{k}) \label{CCaad2}\ee
by defining 
\be a_\lambda(\vec{q}) \equiv  \frac1{\sqrt{c_\alpha(q)}} \alpha_\lambda(\vec{q}), \qquad b_\lambda(\vec{q}) \equiv \frac{\mathcal{F}}{\sqrt{c_\alpha(q)}} \beta_\lambda(\vec{q})
 \ee
 which also leads to properly normalized modes 
 \bea  y_2(\eta, q) &=&   \frac{\sqrt{2} H}{\bp q^{3/2}\sqrt{1+2\frac{H^2}{M_2^2}}}(1+i q\eta) e^{-i q\eta}, \label{GravitonMode} \\ g_2(\eta, q) &=&   \frac{\sqrt{\pi} He^{i\frac{\pi}4\left(1+\sqrt{1-4\frac{M^2_2}{H^2}}\right) }}{\bp q^{3/2}\sqrt{1+2\frac{H^2}{M_2^2}}} (-q\eta)^{3/2}\left(J_{\sqrt{\frac14 -\frac{M_2^2}{H^2}}}(-q\eta) + i Y_{\sqrt{\frac14 -\frac{M_2^2}{H^2}}}(-q\eta) \right) \nonumber \eea
 such that the initial function can be expressed  as follows
\be h_\lambda(\eta, \vec{q}) = a_\lambda(\vec{q}) y_2(\eta, q) +b_\lambda(\vec{q}) g_2(\eta, q) + a^\dagger _{-\lambda}(-\vec{q}) y^*_2(\eta, q)+ b^\dagger _{-\lambda}(-\vec{q}) g^*_2(\eta, q). \label{hDecom}\ee

The commutator on the left of (\ref{CCaad2}) is the standard one for the graviton while the one on the right corresponds to the ghost. In appendix \ref{Tensor perturbations app} we give a rationale for identifying $a_\lambda$ and $b_\lambda$ as annihilation operators and $a_\lambda^\dagger$ and $b_\lambda^\dagger$ as creation operators; we also show there that the Hamiltonian does not have any negative energy eigenvalue provided that the norms of the states created by $a^\dagger_\lambda$ are positive and those with an odd number of $b$-quanta are negative.  Notice that the expression for $y_2$ we find reduces to the standard graviton de Sitter mode when $M_2 \gg H$, but for $H\ll  M_2$ it is suppressed. 
The expressions of $y_2$ and $g_2$ agree with the previous determinations of refs. \cite{Clunan:2009er,Deruelle:2012xv}.
% while the expression of the ghost mode $g_2$ we find agrees up to the form of the exponential 
%\be e^{i\frac{\pi}8\left(\sqrt{1-4\frac{M^2_2}{H^2}}-\sqrt{1-4\frac{M^2_2}{H^2}}^*\right) } 
%= \left\{ \ba {c} e^{i\frac{\pi}4\sqrt{1-4\frac{M^2_2}{H^2}}},
% \quad \mbox{for} \quad M_2>2 H\\  \\1, \quad \mbox{for} \quad M_2<2 H \ea \right.
%\label{expCR}\ee 
%While the case $M_2>2H$ agrees with previous determination, the simplification for $M_2<2 H$ was previously missed. 
However, we observe here that for  $M_2>2H$ a complex exponential  appearing in $g_2$ becomes real,
\be e^{i\frac{\pi}4\sqrt{1-4\frac{M^2_2}{H^2}} }  = e^{-\frac{\pi}4\sqrt{4\frac{M^2_2}{H^2}-1} }, \qquad  (M_2>2H), \ee 
and exponentially suppresses the ghost mode $g_2$ for $M_2 \gg H$ (just like what happened for the vector mode $g_1$). This is what we expect because for $M_2 \gg H$ the effect of the ghost on the inflationary perturbations should disappear.

 Notice that the tensor modes associated with the ghost, $g_2(\eta, q)$, vanish at superhorizon scales, $\eta \rightarrow 0$, even faster than the vector modes. On the other hand, the modes $y_2(\eta, q)$, associated with the ordinary graviton, do not vanish in this limit.

\section{Observational quantities}\label{Observational quantities}

We  consider the power spectrum of the perturbations  that survive at superhorizon scales, namely the curvature perturbation and the perturbations associated with the ordinary graviton, with modes given in eq. (\ref{GravitonMode}), and, if $M_2^2/H^2 \lesssim 1/N$, the extra scalar perturbation $B$. Besides these ons, as usual, the presence of several scalar fields might lead to additional isocurvature modes, which are constrained by observations (see e.g. \cite{Ade:2015lrj}). We do not enter the analysis of such effects because they are  model-dependent and we wish to keep the analysis general here. Such perturbations can be suppressed by an inflationary attractor that effectively reduces the system to a single-field one. The presence of such an attractor has been established in \cite{Kannike:2015apa} in some models of the sort we analyse here  (see also the analysis of the specific model of section \ref{An example: the Higgs and the Higgs of gravity}).

In section \ref{Scalar perturbations} we have seen that the curvature perturbation $\mathcal{R}$ does not receive sizable corrections from the ghost in the slow-roll approximation. Therefore, the associated power spectrum $P_\mathcal{R}(k)$ is (see e.g.  \cite{Sasaki:1995aw})
\be P_\mathcal{R}(q)=\left(\frac{H}{2\pi}\right)^2 N_{,i}N^{,i}.\label{power-spectrum}\ee
 where the field dependent number of e-folds $N(\phi)$ is defined in eq. (\ref{Ndef2})
and in this section we compute the power spectra  at horizon exit $q=a H$.
% and use
The corresponding spectral index $n_s$ is  given in terms of the slow-roll parameters $\epsilon$ and $\eta^i_{\,\,\, j}$ in (\ref{1st-slow-roll}) and (\ref{2nd-slow-roll}) by \cite{Chiba:2008rp,Sasaki:1995aw}
\be  n_s =1-2\epsilon - \frac{ 2 }{ \bar M_{\rm Pl}^2 N_{,i}N^{,i}}+\frac{2\eta_{ij}N^{,i}N^{,j}}{N_{,k}N^{,k}}.  \label{nsFormula}
\ee

In section \ref{Tensor perturbations}, we have seen that the perturbation associated with the ordinary graviton is suppressed with respect to the usual expression in Einstein gravity (see for example the textbook \cite{Liddle:2000cg})  by a factor of $(1+2H^2/M_2^2)^{-1/2}$. Therefore, the power spectrum of tensor perturbations is given by
\be  P_t = \frac{1}{1+\frac{2 H^2}{M_2^2}} \frac{8}{\bar M_{\rm Pl}^2} \left(\frac{H}{2\pi}\right)^2. \label{Ptspectrum} \ee 
By taking the ratio between this equation and  (\ref{power-spectrum}) we obtain the tensor-to-scalar ratio
\be r\equiv \frac{P_t}{P_\mathcal{R}}= \frac{1}{1+\frac{2 H^2}{M_2^2}}\frac{8}{ \bar M_{\rm Pl}^2 N_{,i}N^{,i}}. \label{rW}\ee

Also, we have seen in section \ref{Scalar perturbations} that, in addition to $\mathcal{R}$, there is another scalar perturbation, denoted with $B$, that survives at superhorizon scales for $M_2^2/H^2 \lesssim 1/N$. The power spectrum $P_B$ of the spatial gradient of $B$ has been computed in \cite{Ivanov:2016hcm} for a single-field inflationary model. The results of section \ref{Scalar perturbations} show that the same formula holds for a general matter sector. It turns out to be  the same as the tensor power spectrum in Einstein's gravity,  except that it is smaller by a factor of about $\approx 5$:
\be P_B =\frac{3}{2\bp^2} \left(\frac{H}{2\pi}\right)^2.  \label{PBspectrum} \ee 
 We will conveniently parameterise the effect of  $B$ as it is done for the tensor perturbations: we introduce the ratio 
 \be r' \equiv \frac{P_B }{P_\mathcal{R}}=\frac{3}{2 \bar M_{\rm Pl}^2 N_{,i}N^{,i}}. \label{rp}  \ee
Many models of inflation based on Einstein's gravity predict a tensor power spectrum that is small compared to the curvature power spectrum. Notice also that the correlation between the power spectrum of this isocurvature mode and $\mathcal{R}$ is  suppressed at superhorizon scales: indeed, $B$ contains only the creation and annihilation operators of $\Psi$, while $\mathcal{R}$ only those of the scalar field fluctuations $\varphi^i$ at these scales and in the slow-roll approximation. Therefore, the bounds on isocurvature power spectra of the last Planck data \cite{Ade:2015lrj} can be easily fulfilled.

 We see that the main observational implication of the presence of the ghost in the inflationary perturbations is to suppress $r$ and to introduce another scalar perturbation for small ghost masses, $M_2\lesssim H/\sqrt{N}$. The spectral index $n_s$ and the power spectrum $P_\mathcal{R}$ in general is insensitive to the ghost. We also stress that this conclusion is independent of the matter content of the theory.

 \section{An example: the Higgs and the planckion}\label{An example: the Higgs and the Higgs of gravity}
 
 We apply now the results we obtained to a simple, yet realistic setup: we assume that the only scalar fields that can be active during inflation are the Higgs field, a scalar $s$ that generates the Planck scale\footnote{The conditions on the field content to achieve this dynamical generation of the Planck scale  have been discussed in \cite{Salvio:2014soa}. Here $\xi_s$ is the non minimal coupling between $s$ and the Ricci scalar, appearing in the Lagrangian as $-\xi_s s^2 R/2$.} through its VEV $\langle s\rangle$, that is  $\bp^2 = \xi_s \langle s\rangle^2$, and of course $\zeta$, which corresponds to the $R^2$ term in the Lagrangian (see section \ref{Einstein frame}). Because the Planck mass is due to $s$, this field can be thought of as a Higgs of gravity. In Refs. \cite{Kannike:2015apa,Salvio:2015kka,Salvio:2015jgu} it has been shown that ordinary Higgs inflation \cite{Bezrukov:2007ep,Bezrukov:2008ej,Bezrukov:2009db,Barvinsky:2009ii,Allison:2013uaa,Salvio:2013rja,Salvio:2015cja} (because of the sizable running of its quartic self-coupling \cite{Bezrukov:2012sa,Degrassi:2012ry,Buttazzo:2013uya}) always plays a subdominant role during inflation. One can therefore restrict the attention to $s$ and $\zeta$. 
 
 There is a mixing between $s$ and $\zeta$ and the mass eigenvalues are  \cite{Salvio:2014soa,Kannike:2015apa}
 \be M_{\pm} = \frac{M_s^2 + \xi_c M_0^2}{2} \pm \frac12 \sqrt{(M_s^2 + \xi_c M_0^2)^2 - 4 M_s^2 M_0^2},  \ee
 where $M_0^2\equiv f_0^2 \bp^2/2$, $M_s^2\equiv \langle \partial^2 V/ \partial s^2 \rangle$ and $\xi_c \equiv 1+6\xi_s$. 
 
 In Refs. \cite{Salvio:2014soa,Kannike:2015apa} inflation has been considered in this setup, but assuming that the ghost does not play a significant role. This is always  the case at the background level as the FRW metric is conformally flat and the effect of the Weyl-squared term vanishes. That is also the case  at the linear level in the perturbations whenever  the ghost mass satisfies $M_2 \gg  H$. Here we would like to extend this analysis to smaller values of $M_2$.  By taking into account the Weyl-squared term, in addition  to $\mathcal{R}$  and the tensor perturbation there is in general another relevant perturbation: the isocurvature scalar mode  $B$. The corresponding power spectra have been given in the general case in section \ref{Observational quantities}. We assume here $M_2^2/H^2 \lesssim 1/N$ because the other case $M_2^2/H^2 \gtrsim 1/N$ can be simply obtained by neglecting  $B$. 
 
 As shown in \cite{Kannike:2015apa}, there is an inflationary attractor that effectively reduces the system to a single-field inflationary model: this single field is in general a combination of $s$ and $\zeta$, which, however, reduces to $s$ when $M_0 \gg M_s$ and to $z$ in the other limit. These two cases correspond to the following predictions. 
 \begin{itemize}
 \item   For $M_0 \gg M_s$ ($s$-inflation) the inflationary predictions are
\be n_s\approx 1-\frac{2}{N}\stackrel{N\approx 60}{\approx}0.967,\qquad
r'\approx \frac{3}{2N}\stackrel{N\approx 60}{\approx} 0.025 \qquad
r\approx \frac{1}{1+\frac{2 H^2}{M_2^2}}\frac{8}{N}\stackrel{N\approx 60}{\approx} \frac{0.13}{1+\frac{2 H^2}{M_2^2}} 
%\qquad  A_s \approx \frac{g^4 N^2}{24 \pi^2 \xi_S (1+6\xi_S)}
. \ee
The scalar amplitude $P_R=M_s^2 N^2/(6\pi^2 \bp^2)$
is reproduced for  $M_s\approx 1.4 \times 10^{13}$ GeV.
\item  In the opposite limit ($\zeta$-inflation) one realizes Starobinsky's inflation and the inflationary predictions are
 \be n_s \approx 1 - \frac{2}{N}\stackrel{N\approx 60}{\approx}  0.967,\quad r' \approx \frac{9}{4N^2}\stackrel{N\approx 60}{\approx}  0.0006, \quad r\approx \frac{1}{1+\frac{2 H^2}{M_2^2}}\frac{12}{N^2}\stackrel{N\approx 60}{\approx} \frac{0.003}{1+\frac{2 H^2}{M_2^2}}.\ee
The scalar amplitude $P_R = f_0^2 N^2/(48\pi^2)$
is reproduced for $f_0 \approx 1.8 \times 10^{-5}$.
 \end{itemize}
 \begin{figure}[t]
\begin{center}
  \includegraphics[scale=0.64]{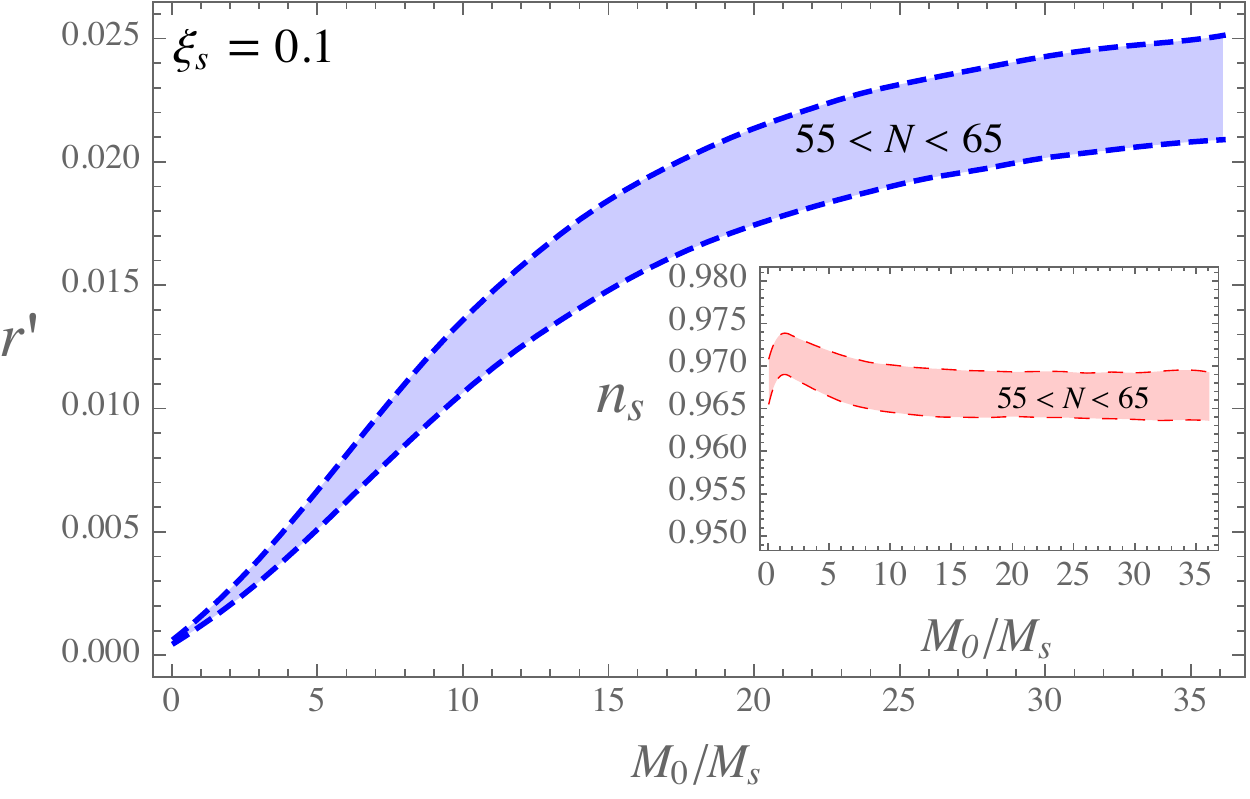}  \hspace{1cm}    \includegraphics[scale=0.64]{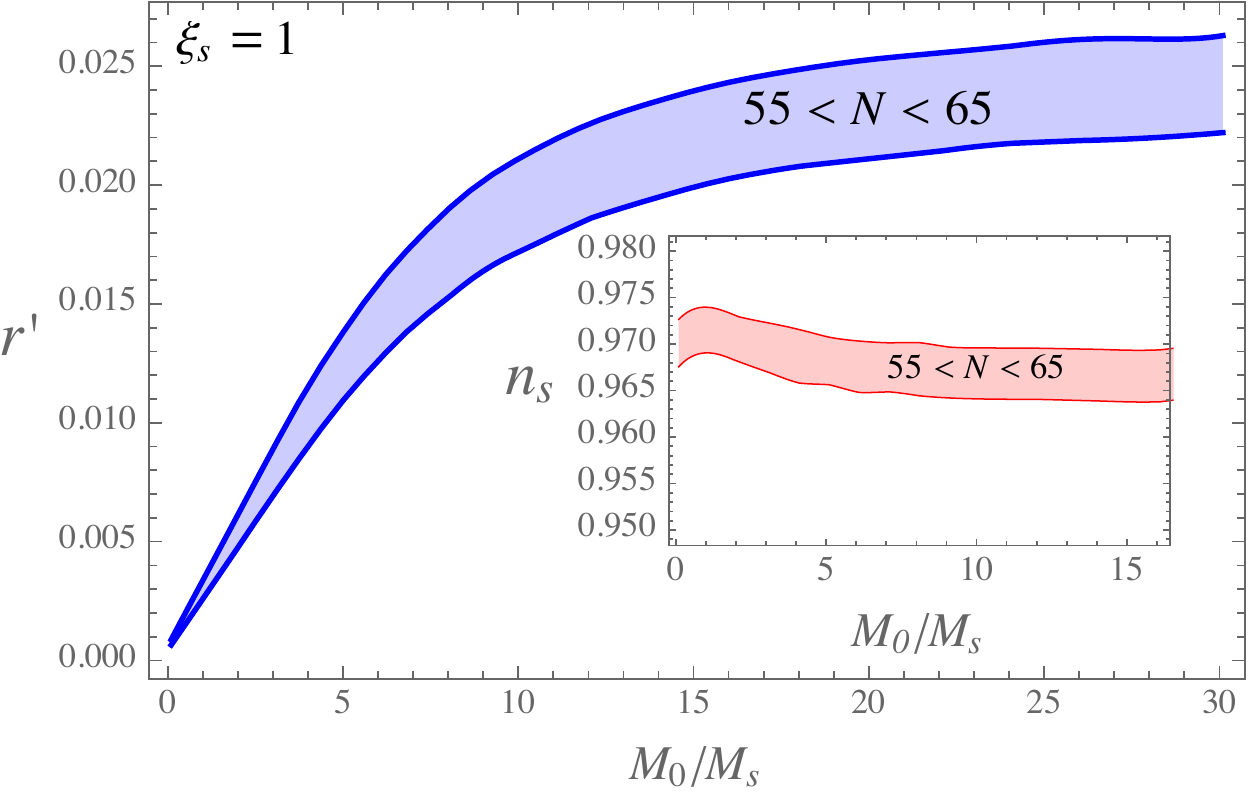} 
  \end{center}
   \caption{\em  The ratio $r'$ between the isocurvature power spectrum $P_B$ of (the gradient of) $B$ and the the curvature power spectrum $P_{\mathcal{R}}$ as defined in eq. (\ref{rp}), as a function of the ratio between the two mass parameters of $\zeta$ and $s$, respectively, setting the dominant inflaton (as explained in the text). The corresponding values of $r$ are given in eq. (\ref{rpTor}). The insets show the values of $n_s$.  An interval of e-folds of $55 < N < 65$ is considered ($N$ suppresses $r'$ and enhances $n_s$). The plot on the left show the results for $\xi_s = 0.1$, while that on the right for $\xi_s =1$.}
\label{rpM0Ms}
\end{figure}

 In fig. \ref{rpM0Ms} we present  $r'$ and $n_s$ for intermediate values of $M_0/M_s$. The corresponding value of $r$ is given by
 \be r = \frac{16 r'}{3(1+\frac{2H^2}{M_2^2})}, \label{rpTor} \ee
 where we  combined (\ref{PBspectrum}), (\ref{rp}), (\ref{Ptspectrum}) and (\ref{rW}). We see that $r$
 is strongly suppressed when $M_2 \ll H$: this limit is relevant because, for example, the maximal value of $M_2$ compatible with  the solution of the hierarchy problem discussed in  \cite{Salvio:2014soa} is $\sim 10^{11}$ GeV, while the typical value of the inflationary  $H$ in this setup is $\sim 10^{13}$ GeV. We find that $r'$ satisfies the bound on isocurvature power spectra of \cite{Ade:2015lrj} by taking an appropriate value of  $f_0$ (this condition is required to match the observed $P_{\mathcal{R}}$ and leads typically to  $f_0\sim10^{-5}$); in the limit of pure $s$-inflation we find, however, that the bounds are very close to the predictions, which suggests that this possibility can be tested with future observations \cite{Finelli:2016cyd}. The prediction of $n_s$ is quite stable as a function of $M_0/M_s$ and agrees well with the bounds of \cite{Ade:2015lrj}. Finally, $r$ always satisfies the bounds of \cite{Ade:2015lrj} when $M_0\ll M_s$. In the opposite limit there is some tension when $M_2  \gtrsim H$ as $s$-inflation is essentially due to a quadratic potential that predicts a rather large value of $r$. This tension, however, disappears if we take instead $M_2 \ll H$, as suggested by  ref.  \cite{Salvio:2014soa}. We see that a relatively light ghost can allow even an inflation due to a quadratic potential.

\section{Issues due to the Weyl-squared term}\label{Issues due to the Weyl term}

  Depending on how the probability is defined, the decay width of the ghost, $\Gamma_2$, may be negative leading to interpretational issues. Feynman has pointed out that negative probabilities are still acceptable if the corresponding events are somewhat unobservables \cite{Feynman}. The  processes affected by $\Gamma_2<0$ in our case  might be interpreted as nothing but intermediate processes of complete events with positive total probability, along the lines of \cite{Lee:1969fy, Feynman}. In the early universe these consist typically of decays of the inflaton into ghosts and other particles followed by ghost decays, which lead to total positive probabilities. 
  
   Ref. \cite{Grinstein:2008bg} pointed out\footnote{See also \cite{Coleman} for a previous discussion by Coleman.}, however, that such intermediate processes  lead to a microscopic violation of causality; to show this the authors considered  two stable particles prepared in some initial state, that come close enough to interact  through the exchange of a ghost, and are later detected. Ref. \cite{Grinstein:2008bg} considered a scalar ghost on flat space; it is not known whether  the same result holds for a spin-2 state on de Sitter space-time and we  leave a complete analysis for future work. However, even if it were the case some necessary conditions should be met to conclude that causality is violated: first, the energy should not be much smaller than the ghost mass in order  to see these effects; second, one should be able to tell where the initial and final particles are and what  their momenta are (otherwise it would not be possible to reconstruct where and when the ghost is annihilated and produced). The first condition forces $H$ or $T$ to be comparable or larger than the ghost mass. The second condition implies that the initial and final particles should be non-relativistic and can only be met if $H$ and the temperature $T$ are much smaller than the mass of the colliding particles.   We observe that stable particles with masses fulfilling the second condition do not necessarily exist in a given no-scale model; for example, in the case of section \ref{An example: the Higgs and the Higgs of gravity} the inflatons have typically masses of order $10^{13}$ GeV, but they are unstable \cite{Kannike:2015apa}. Even if all the necessary conditions to have acausality were satisfied  we now  argue that the  possible acausal processes are  diluted by the expansion of the universe.
  
  First ignore finite temperature effects and consider the inflationary period. In this section we denote with $H_{\rm inf}$ the corresponding Hubble rate.
  Given that (for moderate extensions of the Standard Model) $|\Gamma_2| \lesssim M_2^3/\bp^2$ \cite{Salvio:2016vxi}, for\footnote{The case $M_2\gg H_{\rm inf}$ should be viable  because  corresponds to a decoupled ghost during inflation. }  $M_2\lesssim  H_{\rm inf}$ we have that $\Gamma_2$ is small compared to the Hubble rate during inflation.
  %, which in our case is typically of the order of magnitude of Starobinksy's inflation ($H_{\rm inf} \sim 10^{13}$ GeV) \cite{Kannike:2015apa}
   Indeed, the observational constraint on $H_{\rm inf}$ from the Planck observatory, $H_{\rm inf}<3.6\times10^{-5} \bp$ (see sec 5.1 of  \cite{Ade:2015lrj}) implies $\Gamma_2 \lesssim 
  10^{-9} H_{\rm inf}$. Therefore, the effect of the possible acausal processes   would be diluted by the universe expansion. The dilution takes place even later, as long as $|\Gamma_2| \lesssim  H$, and is described by the Boltzmann equation for the ghost number density $n_g$, that is
 $\dot n_g +3H  n_g \sim - \Gamma_2 n_g,$
which leads to a decreasing $n_g$ for $H \gtrsim |\Gamma_2|$. 
 
  On the other hand, when  $H$ becomes smaller than $\Gamma_2$ it is also much smaller than the ghost mass and the possible acausal processes cannot be observable as argued before (we are using here  $M_2\gg \Gamma_2$  which is amply satisfied whenever\footnote{The opposite case $M_2\gtrsim\bp$ obviously does not create phenomenological problems because corresponds to a ghost decoupled for energies below the Planck scale.} $M_2\ll\bp$). 
  
  Finally, let us consider the finite temperature effects. Effectively, the maximal temperature reached during the universe expansion is the reheating temperature. This quantity has been computed in  \cite{Kannike:2015apa},  at least in some realizations of the no-scale scenario, and turns out to be not larger than  $10^9$ GeV. This is  much smaller than $M_2$ if one saturates the bound $M_2 \lesssim 10^{11}$ GeV of  \cite{Salvio:2014soa}, required to solve the hierarchy problem, and no sizable effects due to the ghost are expected.

\section{Conclusions}

In this work we have analysed all inflationary perturbations in the most general (classically) scale-invariant theory: this includes all terms quadratic in curvature (the $R^2$ and the Weyl-squared terms), the most general no-scale matter Lagrangian and the non-minimal couplings between the scalar fields and $R$. The scales we observe in nature are generated dynamically through dimensional transmutation, in which scale-invariance is broken by quantum effects. 

The main results we have found are the following.

\begin{itemize}
\item We have performed a detailed and careful analysis of all sectors: scalar, vector and tensor perturbations. The corresponding modes are found by means of a Lagrangian approach. We have also shown that the full conserved Hamiltonian for all the perturbations does not feature negative energies if appropriately quantized. An explanation is provided for how the behaviour of ordinary Einstein's gravity coupled to a generic matter sector is recovered when the ghost mass $M_2$ is much bigger than the Hubble rate during inflation $H_{\rm inf}$.

\item The expressions of all the (potentially) observable quantities derived from the relevant power spectra  are presented for the most general scale-invariant theory: the curvature power-spectrum with the corresponding scalar spectra index $n_s$, the tensor power spectrum and the power spectrum of an isocurvature mode $B$ associated  with the helicity-0 component of the spin-2 ghost.

\item Then, these  general results have been applied to a specific concrete model where the  scalar sector features the planckion $s$ (the scalar field whose VEV generates the Planck scale) and the Higgs field, in addition to, of course,  the Starobinsky scalar $\zeta$ due to the $R^2$ term. When $M_2 \gg H_{\rm inf}$ we recover the results of \cite{Kannike:2015apa}. For smaller values of $M_2$ the tensor-to-scalar ratio $r$ becomes suppressed and allows $s$-inflation (which is instead generically in tension with observations for $M_2 > H_{\rm inf}$). More generally, $M_2< H_{\rm inf}$ render viable a large class of models which were in tension with the most recent observations by Planck \cite{Ade:2015lrj}, such as inflationary models with quadratic potentials. For these small values of $M_2$ there is a scalar isocurvature mode, $B$, which, however, is consistent with the bounds on the isocurvature power spectra of \cite{Ade:2015lrj}. Interestingly, however, its power spectrum is rather close to the observational bounds for $s$-inflation, leading to the possibility to test this model with future observations.

\item We have also argued that the possible issues due to the spin-2 ghost associated with the Weyl-squared term do not create phenomenological problems in some no-scale models, at least when $M_2$ is close to the upper bound which ensures the naturalness of the Higgs mass, $M_2 \leq 10^{11}$ GeV.
\end{itemize}

The present work has several possible future applications. For example, it would be interesting to apply the general formul$\ae$ derived here to no-scale models where scale-invariance is broken by non-perturbative effects \cite{Adler:1982ri} and in general to many no-scale models, other than the one considered in section \ref{An example: the Higgs and the Higgs of gravity}. Also, the identification of the quantum perturbations we have performed, with the associated Hilbert space and energy spectra, opens the way to a consistent analysis of non-linear quantum effects on cosmological backgrounds. Perhaps the results of \cite{Salvio:2015gsi} on the quantization of interacting four-derivative theories can be useful in such analysis. 

%\newpage 
\vspace{0.6cm}
\noindent {\bf Acknowledgments.} We thank G. D'amico, C. Germani, A. Kehagias, D. L. Lopez Nacir,  M. M. Ivanov,  A. Strumia and Alexander Vikman for very useful discussions. This work was supported by the grant 669668 -- NEO-NAT -- ERC-AdG-2014.

%\newpage
\appendix

\section{Lagrange and Hamilton methods in four-derivative theories}\label{Lagrange and Hamilton}

Let us consider a physical system described by a certain number of coordinates $q_i$. The case of fields can be obtained by promoting the index $i$ to a space coordinate $\vec{x}$. We restrict our attention to systems described by equations of motion with at most four time-derivatives and with a Lagrangian that depends on $q$, $q'$, $q''$ (we here understand the index $i$) and a possible explicit dependence on time $\eta$:
\be L(q, q', q'', \eta) \ee 
The case in which $L$ does not explicitly depend on time has been considered in \cite{Pais}. We here extend this formalism to a possible time-dependence, which is relevant when studying the perturbations around cosmological backgrounds (see e.g. section \ref{Tensor perturbations}).

The minimal action principle tells us that the variation $\delta S$ of the corresponding action with respect to variations $\delta q$ of the coordinates that vanish on the time boundaries should be zero\footnote{The summation on the index $i$ is understood: for example $\frac{\partial L}{\partial q} \delta q \equiv \sum_i \frac{\partial L}{\partial q_i} \delta q_i$.}: 
\be 0 = \delta S = \int \left(\frac{\partial L}{\partial q} \delta q + \frac{\partial L}{\partial q'} \delta q' +\frac{\partial L}{\partial q''} \delta q''\right) \ee
By integrating by parts the second term once and the third term twice we obtain the Lagrange equations of motion for four-derivative theories
\be \frac{d}{d\eta} \left(\frac{\partial L}{\partial q'} - \frac{d}{d\eta} \frac{\partial L}{\partial q''} \right) = \frac{\partial L}{\partial q}. \ee 

We can move to the Hamilton approach by defining two canonical coordinates 
\be q_1 \equiv q, \qquad q_2 \equiv q'. \label{qlDef} \ee
In this case the useful way of defining the conjugate momenta is 
\be p_l \equiv \frac{\delta  L}{\delta q'_l} \equiv \frac{\partial  L}{\partial q_l'} -\frac{d}{d\eta}\frac{\partial L}{\partial q''_l}, \label{plDef} \ee
where the index $l$ runs over $1,2$. 
Let us see why. First, we assume that we can invert the relations (\ref{qlDef}) and (\ref{plDef}) and obtain $q$, $q'$, $q''$ as a function of $q_1$, $q_2$, $p_1$, $p_2$. This assumption is satisfied for example in the tensor sector studied in section \ref{Tensor perturbations}. Then, define as usual the Hamiltonian $H$ as 
\be H = q_l' p_l - L(q, q', q'', \eta) \label{Hdef}\ee
and regard $H$ as a function of $q_l$, $p_l$ and $\eta$ only:
\be H = H(q_l, p_l, \eta).  \label{Hdep}\ee
We now consider an infinitesimal variation of the Hamiltonian and we compute it in two different ways, by using (\ref{Hdef}) and (\ref{Hdep}). Respectively we have 
\bea dH &=&   p_l dq_l' + q_l' dp_l -  \frac{\partial L}{\partial q} d q - \frac{\partial L}{\partial q'} d q' -\frac{\partial L}{\partial q''}d q'' - \frac{\partial L}{\partial \eta} ,  \\ 
dH &=& \frac{\partial H}{\partial q_l} d q_l + \frac{\partial H}{\partial p_l} d p_l +  \frac{\partial H}{\partial \eta}.  \label{dH2}\eea
By using 
\be \frac{\partial L}{\partial q''} = \frac{\partial L}{\partial q_2'}=\frac{\partial L}{\partial q_2'} -\frac{d}{d\eta} \frac{\partial L}{\partial q_2''} = p_2, \ee
(where we observed that $L$ does not depend on $q_2'' = q'''$ in four-derivative systems)  and 
\be  \frac{\partial L}{\partial q'} = p_1 +\frac{d}{d\eta}  \frac{\partial L}{\partial q''}  \ee
in the first expression of $dH$ we obtain
\be dH = q_l' dp_l  -  \frac{\partial L}{\partial q} d q -\frac{d}{d\eta}  \frac{\partial L}{\partial q''} dq' - \frac{\partial L}{\partial \eta}=q_l' dp_l  -  \frac{\partial L}{\partial q} d q - p_2' dq' - \frac{\partial L}{\partial \eta}. \ee
The use of the Lagrange equation allows to write the term $ \frac{\partial L}{\partial q} d q$ as follows 
\be  \frac{\partial L}{\partial q} d q  =   \frac{d}{d\eta} \left(\frac{\partial L}{\partial q'} - \frac{d}{d\eta} \frac{\partial L}{\partial q''} \right) dq = p_1' dq \ee
so 
\be dH = q_l' dp_l  - p_l' dq_l  - \frac{\partial L}{\partial t}.  \ee 
By comparing now this expression with the one in (\ref{dH2}) we obtain 
\be q_l' = \frac{\partial H}{\partial p_l}, \qquad p_l' = -  \frac{\partial H}{\partial q_l}, \qquad   \frac{\partial H}{\partial \eta}= -\frac{\partial L}{\partial \eta}.  \ee

Therefore we see that in four-derivative systems with a possible time-dependence the Hamilton equations have the standard form provided that the definition of the conjugate momenta are modified according to (\ref{plDef}).

\section{Hamiltonian approach} \label{Hamiltonian approach}

We study here the Hamiltonian of the system on de Sitter space-time and its spectrum. We show that, with an appropriate quantization, the full Hamiltonian has a non-negative spectrum when it is conserved. 

\subsection{Scalar perturbations} 

Let us start with the scalar fields $\varphi$. We will consider this standard case in detail first and then extend the methods developed here to the other sectors. 

 From the expression of the Lagrangian (\ref{LagVarphi}) and the conjugate momentum (\ref{ConjVarphi}) we obtain the Hamiltonian
\be H_{\varphi} = \int d^3x \mathcal{H}_{\varphi} , \qquad \mathcal{H}_{\varphi} = \frac{a^2}{2} \left(\frac{\pi_{\varphi}^2}{a^4} -\varphi \vec{\nabla}^2\varphi + m^2 a^2 \varphi^2  \right). \label{HamiltonS} \ee
In general the Hamiltonian on de Sitter space-time is not conserved, even on the solutions of the equations of motion. However, it becomes conserved at early times, $\eta \rightarrow - \infty$. Notice that in this limit eqs. (\ref{phik-eq}) and (\ref{HamiltonS}) tell us that the effect of the mass, $m$, becomes negligible and we can therefore set $m=0$.  We have
\be H_{\varphi} =2a^2 \int  d^3q \, q^2\left|y_0(\eta,\vec{q}) \right|^2\,\frac12 \left(a_0^\dagger(\vec{q})a_0(\vec{q}) +a_0(\vec{q})a^\dagger_0(\vec{q})\right), \qquad (\eta \rightarrow - \infty)   \ee 
and by using the expression of $y_0$ in (\ref{ModeVarphi}) and $a^2=1/(H\eta)^2$ we find 
\be H_{\varphi} =  \int d^3q  \,   \frac{q}2 \left(a_0^\dagger(\vec{q})a_0(\vec{q}) +a_0(\vec{q})a^\dagger_0(\vec{q})\right), \qquad (\eta \rightarrow - \infty).  \ee
The fact that $a_0^\dagger$ and $a_0$ respectively creates and annihilates quanta of positive norm guarantees that the Hamiltonian has positive eigenvalues. 

If one repeats the same steps for the scalar perturbation $\hat \Psi$ of the metric one finds that its term in the Hamiltonian is 
\be H_{\Psi} = - \int d^3q  \,  \frac{q}2 \left( b_0(\vec{q})^\dagger  b_0(\vec{q}) +b_0(\vec{q})  b_0(\vec{q})^\dagger\right), \qquad (\eta \rightarrow - \infty). \ee
In this case one can ensure that the spectrum of $H_{\Psi}$ is bounded from below by introducing a negative norm for the states with an odd number of quanta created and annihilated by $b_0^\dagger$ and $b_0$ respectively. 

\subsection{Vector perturbations}\label{Vector perturbations app}
 By using the Lagrangian in (\ref{LagV}) and the momenta in (\ref{PiDef}) the  Hamiltonian for vector perturbations turns out to be 
\be H_V = \int d^3x \left[\frac{M_2^2}{\bp^2}P_j  \vec{\nabla}^{-2} P_j+\frac{\bp^2}{4}\left(a^2 V_{j}\vec{\nabla}^2 V_{j}-\frac1{M_2^2} V_j \vec{\nabla}^4 V_j\right)\right]. \ee
Notice that it contains the inverse Laplacian operator $\vec{\nabla}^{-2}$.
This inverse operator can be treated by going to momentum space: by inserting the Fourier decomposition (\ref{ExpMomV}) in   $H_V$ we obtain
\be   H_V =
 -\sum_{\lambda=\pm 1} \int d^3 q \left[\frac{M_2^2}{\bp^2 q^2}(P_\lambda)^\dagger  P_\lambda +\frac{\bp^2q^2}{4} (V_\lambda)^\dagger\left(a^2+\frac{q^2}{M_2^2}\right)V_\lambda\right].\ee
% \xxx{comment on the negative overall sign}. 
By using the equations of motion in (\ref{EqVpert}) again one can show that the Hamiltonian is conserved at early times, $\eta \rightarrow - \infty$. Also, in the same limit, we find 
\be H_{V} = - \sum_{\lambda=\pm 1}\int d^3q   \,  \frac{q}{2}\left(c^\dagger_\lambda({\vec{q}})c_\lambda({\vec{q}})+c_\lambda({\vec{q}})c^\dagger_\lambda({\vec{q}})\right), \qquad (\eta \rightarrow - \infty). \ee
and we can proceed to stabilize the Hamiltonian as we did for the helicity zero field $\hat \Psi$.

\subsection{Tensor perturbations}\label{Tensor perturbations app}

The Hamiltonian for the tensor perturbations is
\bea && H_T =\int d^3 x  \left(p^{(a)}_{ij} h_{ij}'^{(a)}-  \Lag_{T} \right)=
 \int d^3x \left[p^{(1)}_{ij} h_{ij}^{(2)}-\frac{2M_2^2}{\bp^2}  p^{(2)}_{ij}  p^{(2)}_{ij} \right. \nonumber \\ 
&&\left.  -\frac{\bp^2}{8} h_{ij}^{(2)} \left(a^2-\frac{2}{M_2^2} \vec{\nabla}^2 \right)h_{ij}^{(2)} -\frac{\bp^2}{8} h_{ij}^{(1)} \left(a^2\vec{\nabla}^2-\frac{1}{M_2^2} \vec{\nabla}^4 \right)h_{ij}^{(1)}\right].  \label{Htensor} \eea
%\xxx{comments on Dirac-Pauli quantization}
Notice that inserting the expansion (\ref{ExpMom}) in (\ref{Htensor}) leads to
\bea  H_T &=&
 \sum_{\lambda=\pm 2} \int d^3 q  \left[(p^{(1)}_{\lambda})^\dagger h_{\lambda}^{(2)}-\frac{2M_2^2}{\bp^2}  (p^{(2)}_{\lambda})^\dagger  p^{(2)}_{\lambda}  \right.  
\\ \nonumber  
&&  \left.   -\frac{\bp^2}{8} (h_{\lambda}^{(2)})^\dagger \left(a^2+\frac{2 q^2}{M_2^2} \right)h_{\lambda}^{(2)} +\frac{\bp^2q^2}{8} (h_{\lambda}^{(1)})^\dagger \left(a^2+\frac{q^2}{M_2^2} \right)h_{\lambda}^{(1)}\right].  
\label{Hmom}  \eea

By using the field decomposition in (\ref{hDecom}) and the expressions for the  canonical variables in (\ref{canCoord}) and (\ref{CanMom}) we find (in the $\eta \rightarrow -\infty$ limit)
\bea  H_T &=&
 \sum_{\lambda=\pm 2} \int d^3 q  \left\{\frac{q M_2^2}{2 H^2+M_2^2} \left[ a_\lambda^\dagger(\vec{q})a_\lambda(\vec{q}) - \left(1+4 \frac{H^2}{M_2^2}\right) b_\lambda^\dagger(\vec{q})b_\lambda(\vec{q}) \right. \right. \\ \nonumber
&& \hspace{2cm} \left. \left. - \frac{2 i H^2}{M_2^2} ( a_\lambda^\dagger(\vec{q})b_\lambda(\vec{q})- b_\lambda^\dagger(\vec{q})a_\lambda(\vec{q})) \right] +\frac{q}{2} \left([a_\lambda(\vec{q}), a^\dagger_\lambda(\vec{q})] - [b_\lambda(\vec{q}), b^\dagger_\lambda(\vec{q})]   \right)  \right\}.  \eea
This shows that also the tensor Hamiltonian is conserved in this limit, but it is non-obvious whether its spectrum is non-negative. 

In order to see that this is the case we consider the set of transformations
\bea  \tilde a_\lambda &=&\frac1{\sqrt{c}} \left(a_\lambda + i \sqrt{1-c} \, b_\lambda \right), \label{BV1} \\ 
\tilde b_\lambda &=&\sqrt{\frac{1-c}{c}} \left(a_\lambda + i\frac{ b_\lambda}{\sqrt{1-c}} \right),  \label{BV2}  \eea
where $c$ is a real number constrained by $0 < c < 1$, but otherwise arbitrary. Notice that, for any $c$, the commutation rules of $\tilde a_\lambda$ and $\tilde b_\lambda$  are the same as those of $a_\lambda$ and $b_\lambda$, respectively (see eq. (\ref{CCaad2})):  
\be [\tilde a_\lambda(\vec{q}), \tilde a_{\lambda'}^\dagger(\vec{k})] = \delta_{\lambda \lambda'}\delta^{(3)}(\vec{q}-\vec{k}),\qquad  [\tilde b_\lambda(\vec{q}), \tilde b_{\lambda'}^\dagger(\vec{k})] = -   \delta_{\lambda \lambda'}\delta^{(3)}(\vec{q}-\vec{k}). \label{CCaad3}\ee
By inserting the transformation (\ref{BV1})-(\ref{BV2}) in the Hamiltonian $H_T$ one finds an expression that apparently depends on $c$. However, this dependence is spurious because $c$ is only an arbitrary parameter of the transformation and cannot affect the physical properties. In particular we can take the small $c$ limit, where the Hamiltonian simply reduces to 
\be  H_T =
 \sum_{\lambda=\pm 2} \int d^3 q \frac{q}{2} 
 \left[ \tilde a_\lambda^\dagger(\vec{q}) \tilde a_\lambda(\vec{q}) + \tilde a_\lambda(\vec{q}) \tilde a^\dagger_\lambda(\vec{q})
  - \left( \tilde b_\lambda^\dagger(\vec{q}) \tilde b_\lambda(\vec{q}) + \tilde b_\lambda(\vec{q}) \tilde b^\dagger_\lambda(\vec{q})\right)\right]. \label{HTsimple}
 \ee
 Therefore, we see that $H_T$ does not have negative eigenvalues provided that the norms of the states created by $ \tilde a^\dagger_\lambda$ are all positive, while those with an odd number of quanta created by $ \tilde b^\dagger_\lambda$ are negative. It is easy to see that the same conditions  hold for the original operators $a^\dagger_\lambda$ and $b^\dagger_\lambda$ by using the transformation for (\ref{BV1}) and (\ref{BV2}).

 \vspace{2cm}
 
 \footnotesize
\begin{multicols}{2}

\end{multicols}

\end{document}